\documentclass[
  10pt,
  aps,
  pre,
  superscriptaddress, 
  showkeys, 
  twocolumn,
  letterpaper,
]{revtex4-1}

\usepackage{amsmath}
\usepackage[utf8]{inputenc}
\usepackage{amssymb}
\usepackage{bm}
\usepackage[squaren]{SIunits}
\usepackage{graphicx}

\newcommand{\ivec}[1]{{#1}}
\renewcommand{\vec}[1]{{\bm{#1}}}
\newcommand{\itens}[1]{{\mathsf{#1}}}
\newcommand{\tens}[1]{{\bm{\mathsf{#1}}}}
\newcommand{\rot}{R}
\newcommand{\trace}[1]{{\mathrm{Tr}\,{#1}}}

\newcommand{\nem}[1]{\tilde{#1}}
\newcommand{\unem}[1]{\tilde{#1}}
\newcommand{\norm}[1]{\vert#1\vert}
\renewcommand{\d}{{\mathrm{d}}}

\newcommand{\D}{{\mathrm{D}}}
\newcommand{\bondindex}[1]{{\langle#1\rangle}}

\newcommand{\tref}[1]{Table~\ref{#1}}
\newcommand{\fref}[1]{Fig.~\ref{#1}}
\newcommand{\fsref}[1]{Figs.~\ref{#1}}

\newcommand{\Fref}[1]{Fig.~\ref{#1}}
\newcommand{\eref}[1]{Eq.~\eqref{#1}}
\newcommand{\esref}[1]{Eqs.~\eqref{#1}}
\newcommand{\seref}[1]{\eqref{#1}}
\newcommand{\sref}[1]{Section~\ref{#1}}
\newcommand{\ssref}[1]{Sections~\ref{#1}}
\newcommand{\aref}[1]{Appendix~\ref{#1}}
\newcommand{\asref}[1]{Appendices~\ref{#1}}

\newenvironment{myTable}[3]{\begin{table}\caption{\label{#1}#2}\begin{tabular}{#3}\hline}{\hline\end{tabular}\end{table}}
\newcommand{\mr}{\hline}
\newcommand{\printbib}{\bibliography{paper}}

\allowdisplaybreaks

\begin{document}
\title[Triangles bridge the scales]{Triangles bridge the scales: Quantifying cellular contributions to tissue deformation}
\date{\today}
\author{Matthias Merkel}
\email{mmerkel@syr.edu}
\affiliation{Max Planck Institute for the Physics of Complex Systems, Nöthnitzer Str.\ 8, 01187 Dresden, Germany}
\affiliation{Department of Physics, Syracuse University, Syracuse, New York 13244, USA}
\author{Raphaël Etournay}
\affiliation{Max Planck Institute of Molecular Cell Biology and Genetics, Pfotenhauerstr.\ 108, 01307 Dresden, Germany}
\affiliation{Unité de Génétique et Physiologie de l’Audition, Institut Pasteur, 75015 Paris, France}
\author{Marko Popovi\'c}
\affiliation{Max Planck Institute for the Physics of Complex Systems, Nöthnitzer Str.\ 8, 01187 Dresden, Germany}
\author{Guillaume Salbreux}
\affiliation{Max Planck Institute for the Physics of Complex Systems, Nöthnitzer Str.\ 8, 01187 Dresden, Germany}
\affiliation{Francis Crick Institute, Lincoln's Inn Fields Laboratory, 44 Lincoln's Inn Fields, London WC2A~3LY, UK}
\author{Suzanne Eaton}
\affiliation{Max Planck Institute of Molecular Cell Biology and Genetics, Pfotenhauerstr.\ 108, 01307 Dresden, Germany}
\author{Frank Jülicher}
\email{julicher@pks.mpg.de}
\affiliation{Max Planck Institute for the Physics of Complex Systems, Nöthnitzer Str.\ 8, 01187 Dresden, Germany}


\begin{abstract}
In this article, we propose a general framework to study the dynamics and topology of cellular networks that capture the geometry of cell packings in two-dimensional tissues.
Such epithelia undergo large-scale deformation during morphogenesis of a multicellular organism.
Large-scale deformations emerge from many individual cellular events such as cell shape changes, cell rearrangements, cell divisions, and cell extrusions.
Using a triangle-based representation of cellular network geometry, we obtain an exact decomposition of large-scale material deformation. 
Interestingly, our approach reveals contributions of correlations between cellular rotations and elongation as well as cellular growth and elongation to tissue deformation.
Using this Triangle Method, we discuss tissue remodeling in the developing pupal wing of the fly \textit{Drosophila melanogaster}. 
\end{abstract}

\maketitle

\begin{figure*}
  \centering
  \includegraphics{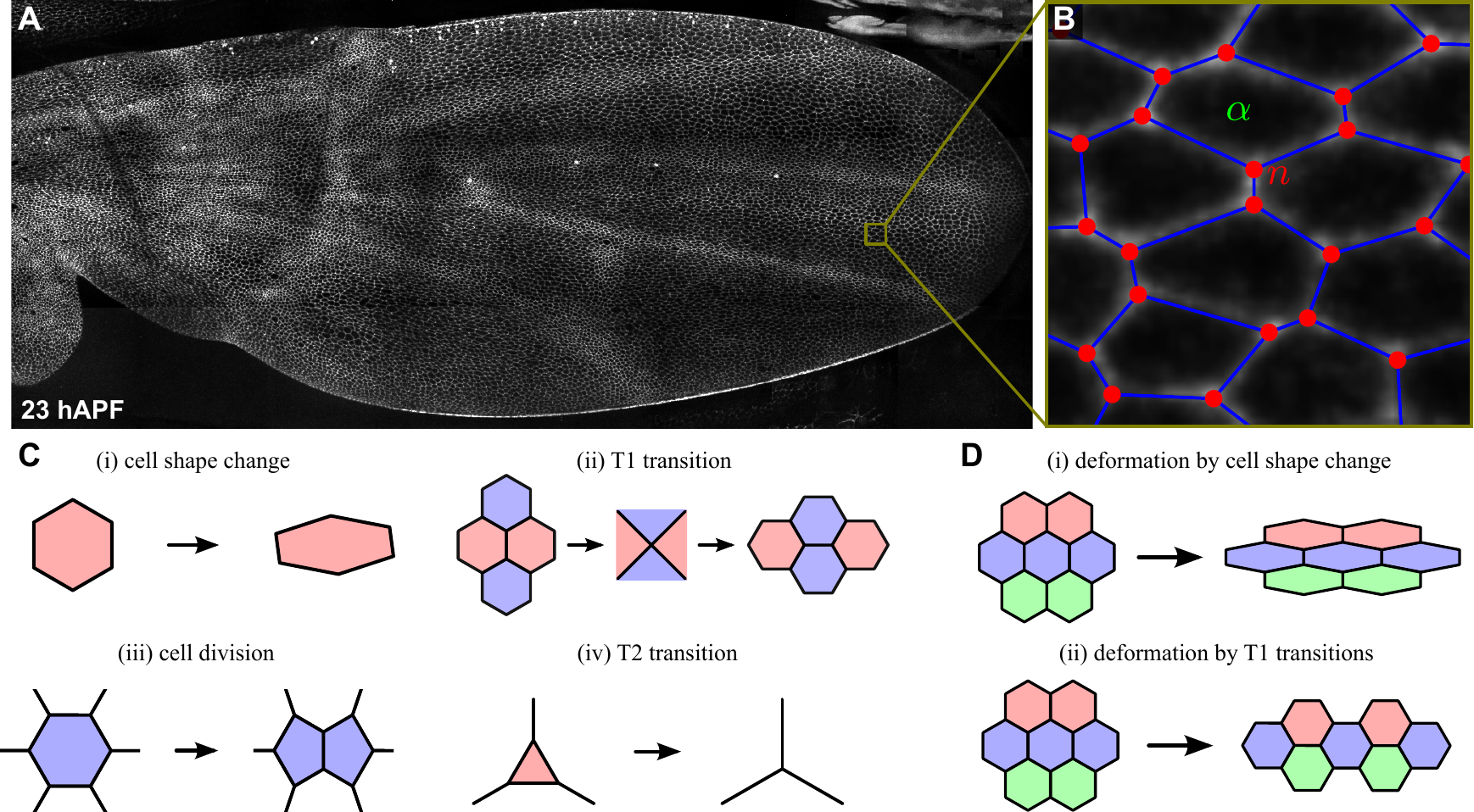}
  \caption{(A) The developing fly wing is an important model system to study epithelial morphogenesis. This panel shows the wing blade at the developmental time of 23 hours after puparium formation (hAPF). (B) Magnified region of membrane-stained wing tissue overlaid with the corresponding polygonal network. Cells are represented by polygons (green), cell-cell interfaces correspond to polygon edges (blue), and polygon corners correspond to vertices (red). (C) We consider four kinds of cell-scale processes. (D) Two examples for pure shear of a piece of cellular material. (i) Pure shear by cell shape change. (ii) Pure shear by T1 transitions. Colors in panels C and D indicate cell identities.\label{fig:introduction}}
\end{figure*}
\tableofcontents

\section{Introduction}
Morphogenesis is the process in which a complex organism forms from a fertilized egg.
Such morphogenesis involves the formation and dynamic reorganization of tissues \cite{Wolpert2001,Blankenship2006,Aigouy2010,Bosveld2012,Merkel2014,Etournay2015}.
An important type of tissues are epithelia, which are composed of two-dimensional layers of cells. 
During development, epithelia can undergo large-scale remodeling and deformations. This tissue dynamics can be driven by both internal and external stresses \cite{Aigouy2010,Etournay2015}. Large-scale deformations are the result of many individual cellular processes such as cellular shape changes, cell divisions, cell rearrangements, and cell extrusions. 
The relationship between cellular processes and large-scale tissue deformations is key for an understanding of morphogenetic processes.
In this paper, we provide a theoretical framework that can exactly relate cellular events to large-scale tissue deformations.

Modern microscopy techniques provide live image data of the development of animal tissues \textit{in vivo} \cite{Keller2008,Aigouy2010,Bosveld2012,Merkel2014,Etournay2015,Etournay2016}. An important example is the fly wing, where about $10^4$ cells have been tracked over $17$ hours (\fref{fig:introduction}A) \cite{Etournay2015}. Using cell membrane markers, semi-automated image analysis can segment the geometrical outlines and the neighbor relationships of all observed cells, and track their lineage throughout the process (\fref{fig:introduction}B) \cite{Wiesmann2015,Aigouy2010,Mosaliganti2012,BarbierdeReuille2015,Cilla2015,Etournay2016}. 
This provides detailed information about many different cellular events such as cell shape changes, cell rearrangements, cell division, and cell extrusions.

As a result of a large number of such cellular events, the cellular network is remodeled and undergoes changes in shape. Such shape changes can be described as tissue deformations using concepts from continuum mechanics. The aim of this paper is to provide a framework to describe the geometry of tissue remodeling at different scales. 
We identify the contributions to tissue deformation stemming from cell shape changes and from distinct cellular processes that remodel the cellular network (\fref{fig:introduction}C). 
For example, tissue shear can result from shape changes of individual cells or alternatively from cell rearrangements without cells changing their shape (\fref{fig:introduction}D).
In general, tissue deformations involve a combination of such events.
Furthermore, cell divisions and extrusions also contribute to tissue deformations.

The relationship between tissue deformations and cellular events have been discussed in previous work \cite{Brodland2006,Graner2008,Blanchard2009,Kabla2010,Economou2013,Guirao2015}. Here, in order to obtain an exact decomposition of tissue deformation, we present a Triangle Method that is based on the dual network to the polygonal cellular network. We have recently presented a quantitative study of the \textit{Drosophila} pupal wing morphogenesis using this approach \cite{Etournay2015}.

In the following sections \ssref{sec:polygonalTriangularNetworks}-\ref{sec:allTogether}, we provide the mathematical foundations of the Triangle Method to characterize tissue remodeling.
In \sref{sec:polygonalTriangularNetworks}, we introduce a polygonal network description of epithelial cell packings. We discuss different types of topological changes of the network that are associated with cellular rearrangements and we define the deformation fields of the network. 
In \sref{sec:triangles}, we define mathematical objects that characterize triangle geometry and derive the relation between triangle shape changes and network deformations. 
\sref{sec:ttcontributionsToDeformation} presents the contribution of individual topological changes to network deformations.
\sref{sec:allTogether} combines the concepts developed in the previous sections. We discuss the decomposition of large-scale tissue deformation in the contributions resulting from large numbers of individual cellular processes.
In \sref{sec:flyWing}, we apply the Triangle Method to the developing fly wing, comparing morphogenetic processes in different subsections of the wing blade. 
Finally, we discuss our results in \sref{sec:discussion}. Technical details are provided in the \asref{app:firstAppendix}--\ref{app:lastAppendix}.

\section{Polygonal and triangular networks}
\label{sec:polygonalTriangularNetworks}
We introduce quantities to characterize small-scale and large-scale material deformation. To this end, we first discuss two complementary descriptions of epithelial cell packing geometry.

\subsection{Description of epithelia as a network of polygons}
\label{sec:polygonalNetworks}
The cell packing geometry of a flat epithelium can be described by a network of polygons, where each cell is represented by a polygon and each cell-cell interface corresponds to a polygon edge (\fref{fig:introduction}B, \fref{fig:triangulation}A)
\footnote{The polygonal network is introduced just for the sake of clarity here. All of our results are equally applicable for a much broader class of cellular networks where cell outlines may be curved.}. Polygon corners are referred to as vertices, and a vertex belonging to $M$ polygons is denoted $M$-fold vertex.
Thus, the polygonal network captures the topology and geometry of the junctional network of the epithelium.

Within such a polygonal network, we consider four kinds of cellular processes (\fref{fig:introduction}C). (i) Polygons may change their shapes due to movement of vertices. (ii) Polygons may rearrange by changing their neighbors. A T1 transition is an elementary neighbor exchange during which two cells (red) lose their common edge, and two other cells (blue) gain a common edge. However, a T1 transition could also just occur partially. For instance, a single edge can shrink to length zero giving rise to an $M$-fold vertex with $M>3$. Conversely, an $M$-fold vertex with $M>3$ can split into two vertices that are connected by an edge. (iii) A polygon may split into two by cell division. (iv) A T2 transition corresponds to the extrusion of a cell from the network such that the corresponding polygon shrinks to a vertex. Note that the first process corresponds to a purely geometrical deformation whereas the last three processes correspond to topological transitions in the cellular network.

\subsection{Triangulation of a polygonal network}
\begin{figure}
  \centering
  \includegraphics{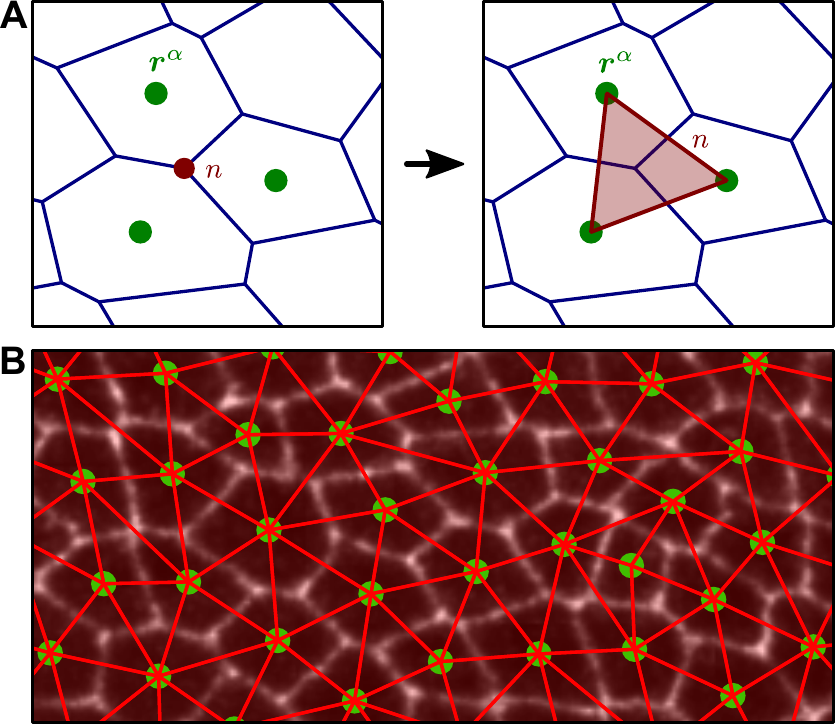}
  \caption{Triangulation of the cellular network. (A) Each three-fold vertex $n$ (red dot) gives rise to a single triangle (red), which is also denoted by $n$. The corners of the triangle are defined by the centers of the three abutting cells (green dots). (B) Triangulation (red) on top of membrane-stained biological tissue (white). There are no gaps between the triangles.\label{fig:triangulation}}
\end{figure}
\begin{myTable}{tab:notation}{Notation used throughout this article.}{p{2.8cm}p{5.5cm}}
Examples &  \\ 
\mr
$\alpha,\beta,\gamma$ & Cell indices \\
$n,m$ & Vertex and triangle indices \\
$i,j,k$ & Dimension indices (either $x$ or $y$)\\\hline
$\vec{r},\vec{h}\;$ and $\;\ivec{r}_i,\ivec{h}_i$ & Vectors\\
$\tens{U},\tens{s}^n\;$ and $\;\itens{U}_{ij},\itens{s}_{ij}^n$ & Tensors \\
$\tens{\unem{U}},\tens{\nem{q}}^n\;$ and $\;\itens{\unem{U}}_{ij},\itens{\nem{q}}_{ij}^n$ & Symmetric, traceless tensors \\\hline
$A,\itens{\nem{Q}}_{ij},\itens{U}_{ij}$ & Large-scale quantities\\
$a^n,\itens{\nem{q}}_{ij}^n,\itens{u}_{ij}^n$ & Triangle-related quantities\\\hline
$\Delta A, \Delta\itens{\nem{Q}}_{ij}$ & Finite quantities\\
$\delta A, \delta\itens{U}_{ij}$ & Infinitesimal quantities\\
\end{myTable}
To define contributions of cellular processes to the large-scale deformation of a polygonal network, we introduce a triangulation of the polygonal network (\fref{fig:triangulation}A).
For each vertex $n$ (red) being surrounded by three cells, a triangle $n$ (red) is created by defining its corners to coincide with the centers $\vec{r}^\alpha$ of the three cells (green).
For the special case of an $M$-fold vertex with $M>3$, we introduce $M$ triangles as described in \aref{app:triangulation}.
The center of a given cell $\alpha$ is defined by the vector
\begin{equation}
  \vec{r}^\alpha = \frac{1}{a^\alpha}\int_{a^\alpha}\vec{r}\,\d A\text{,}
\end{equation}
where the integration is over the cell area $a^\alpha$ and $\vec{r}$ is a position vector (\tref{tab:notation}).
Since triangle corners correspond to cell centers, oriented triangle sides are referred to by a pair of cell indices $\bondindex{\alpha\beta}$, and the corresponding triangle side vector is given by
\begin{equation}
  \vec{r}^\bondindex{\alpha\beta} = \vec{r}^{\beta}-\vec{r}^{\alpha}\text{.}
\end{equation}
The so-created triangulation of the cellular material contains no gaps between the triangles. It can be regarded as the dual of the polygonal network (\fref{fig:triangulation}B).

\subsection{The deformation tensor}
To characterize the deformation of the cellular network, we define a deformation tensor $\itens{U}_{ij}$ that corresponds to the coarse-grained displacement gradient:
\begin{equation}
  \itens{U}_{ij} = \frac{1}{A}\int{\partial_i\ivec{h}_j\,\d A}\text{.}\label{eq:UOmegaContinuousArea}
\end{equation}
Here, $A$ is the area of the coarse-graining region. The vector field $\vec{h}(\vec{r})$ describes the continuous displacement field with respect to the reference position $\vec{r}$, and the indices $i,j$ denote the axes $x,y$ of a Cartesian coordinate system. The region may in general encompass several cells or just parts of a single cell.

The deformation tensor $\itens{U}_{ij}$ can be expressed in terms of the displacements $\vec{h}(\vec{r})$ along the margin of the region (see \aref{app:genGauss}):
\begin{equation}
  \itens{U}_{ij} = \frac{1}{A}\oint{\ivec{h}_j\ivec{\nu}_i\,\d\ell}\text{.}\label{eq:UOmegaContinuousBoundary}
\end{equation}
Here, the vector $\vec{\nu}$ denotes the local unit vector that is normal to the margin pointing outwards. 

\begin{figure}
  \centering
  \includegraphics{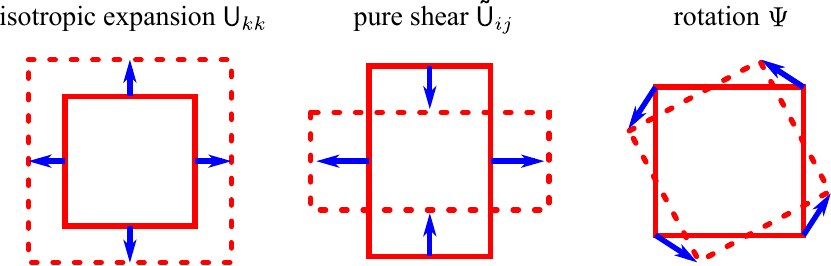}
  \caption{The deformation tensor $\itens{U}_{ij}$ can be decomposed into isotropic expansion, pure shear, and rotation. In particular for $\itens{U}_{ij}\ll 1$, these deformation components correspond to trace $\itens{U}_{kk}$, symmetric, traceless part $\itens{\unem{U}}_{ij}$, and antisymmetric part $\Psi$, respectively.\label{fig:decompositionU}}
\end{figure}
We define trace $\itens{U}_{kk}$, symmetric, traceless part $\itens{\unem{U}}_{ij}$, and antisymmetric part $\Psi$ of the deformation tensor $\itens{U}_{ij}$ as follows:
\begin{equation}
  \itens{U}_{ij} = \frac{1}{2}\itens{U}_{kk}\itens{\delta}_{ij} + \itens{\unem{U}}_{ij} - \Psi\itens{\epsilon}_{ij}\text{.}\label{eq:decompositionU}
\end{equation}
Here, $\itens{\delta}_{ij}$ denotes the Kronecker symbol and $\itens{\epsilon}_{ij}$ is the generator of counter-clockwise rotations with $\itens{\epsilon}_{xy}=-1$, $\itens{\epsilon}_{yx}=1$ and $\itens{\epsilon}_{xx}=\itens{\epsilon}_{yy}=0$. Here and in the following, all symmetric, traceless tensors are marked with a tilde as $\itens{\unem{U}}_{ij}$ is. For small displacement gradients $\itens{U}_{ij}\ll 1$, the components of $\itens{U}_{ij}$ can be respectively interpreted as isotropic expansion $\itens{U}_{kk}$, pure shear $\itens{\unem{U}}_{ij}$, and rotation by the angle $\Psi$ (\fref{fig:decompositionU}).

\esref{eq:UOmegaContinuousArea} and \seref{eq:UOmegaContinuousBoundary} define the deformation tensor $\itens{U}_{ij}$ based on the continuous displacement field $\vec{h}(\vec{r})$. However for typical experiments, the displacement $\vec{h}(\vec{r})$ is only known for a finite number of positions $\vec{r}$. In the following, we will thus focus on the displacements of cell center positions $\vec{h}(\vec{r}^\alpha)=\vec{h}^\alpha$ and interpolate between them in order to compute the deformation tensor $\itens{U}_{ij}$.

\subsection{Triangle-based characterization of network deformation}
\label{sec:triangulation}
\begin{figure}
  \centering
  \includegraphics{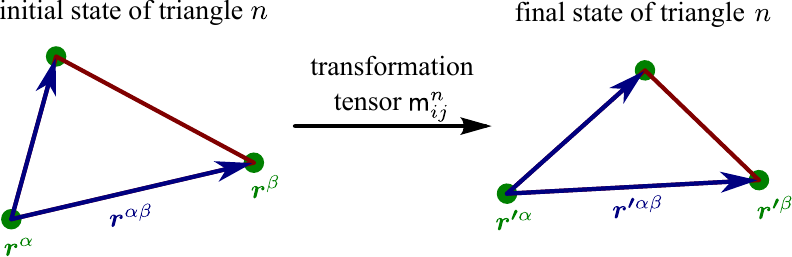}
  \caption{Deformation of a triangle $n$ from an initial state to a final state. The deformation is characterized by the linear transformation tensor $\itens{m}_{ij}^n$ mapping the initial sides vectors of the triangle to the final side vectors (blue arrows).\label{fig:triangleDeformation}}
\end{figure}
We relate the large-scale deformation characterized by $\itens{U}_{ij}$ to small-scale deformation, which we quantify on the single-triangle level. 
We describe the deformation of a single triangle $n$ from an initial to a final state by an affine transformation, which is characterized by a transformation tensor $\itens{m}_{ij}^n$ that maps each initial triangle side vector $\vec{r}^\bondindex{\alpha\beta}$ to the corresponding final side vector $\vec{r}^{\prime\bondindex{\alpha\beta}}$ (\fref{fig:triangleDeformation}):
\begin{equation}
  \ivec{r}_i^{\prime\bondindex{\alpha\beta}} = \itens{m}_{ij}^n\ivec{r}_j^\bondindex{\alpha\beta}\text{.}\label{eq:triangleTransformationTensor}
\end{equation}
Note that \eref{eq:triangleTransformationTensor} uniquely defines the tensor $\itens{m}_{ij}^n$, which always exists \footnote{As long as the initial triangle has nonzero area.}. However for polygons with more than three sides, no such tensor $\itens{m}_{ij}^n$ exists in general. This is the deeper reason for us to choose a triangle-based approach.

\begin{figure*}
  \centering
  \includegraphics{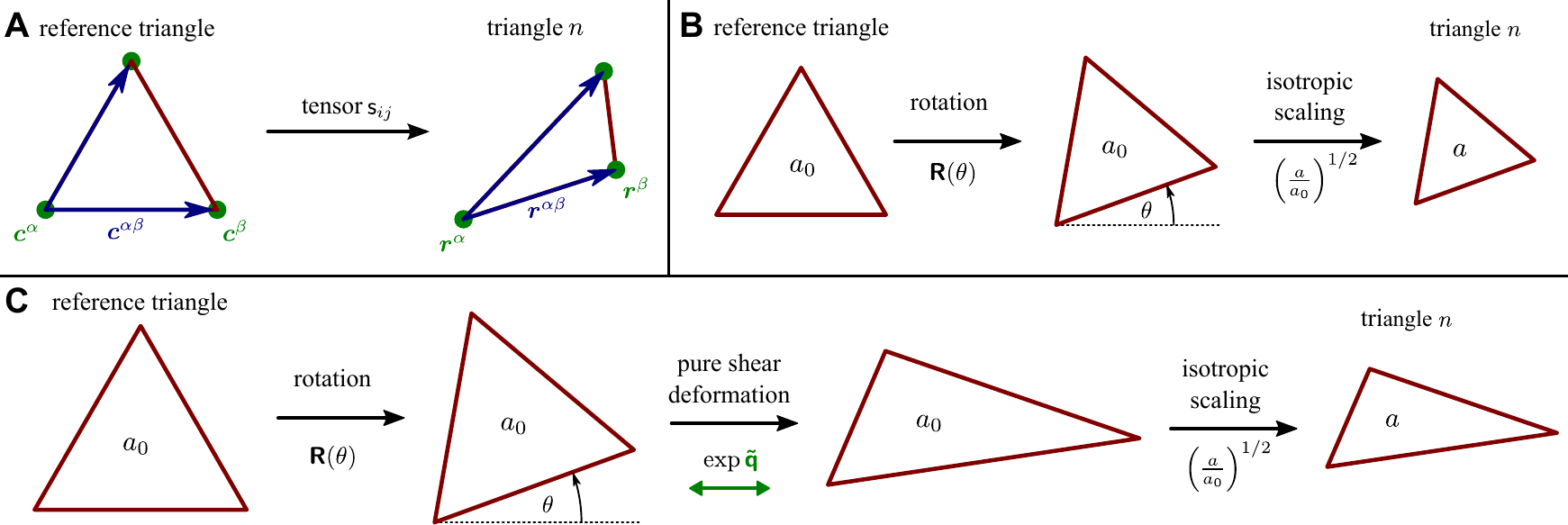}
  \caption{Characterization of triangle shape. 
  (A) The shape of a triangle $n$ in a given state is characterized by the tensor $\itens{s}_{ij}$. Tensor $\itens{s}_{ij}$ maps the side vectors of a virtual equilateral reference triangle to the side vectors of triangle $n$ (blue arrows). 
  (B) Shape properties for the special case of an equilateral triangle $n$. To extract shape properties, the transformation tensor $\itens{s}_{ij}$ is decomposed into a counter-clockwise rotation by a triangle orientation angle $\theta$ and an isotropic rescaling to match the actual triangle area $a$. 
  (C) Shape properties for for the general case of an elongated triangle $n$. The transformation tensor $\itens{s}_{ij}$ is decomposed into a counter-clockwise rotation by the triangle orientation angle $\theta$, a pure shear deformation characterized by the triangle elongation tensor $\itens{\nem{q}}_{ij}$, and an isotropic rescaling to match the actual triangle area $a$.\label{fig:triangleShapeNew}}
\end{figure*}
To relate triangle deformation to large-scale deformation $\itens{U}_{ij}$, we first define a continuous displacement field $\vec{h}(\vec{r})$ by linearly interpolating between cell center displacements $\vec{h}^\alpha$. For any position $\vec{r}$ that lies within a given triangle $n$, we define:
\begin{equation}
  \ivec{h}_j(\vec{r}) = \ivec{h}^\alpha_j + (\ivec{r}_i-\ivec{r}_i^\alpha)\itens{u}_{ij}^n\text{.}\label{eq:triangleDefinitionU}
\end{equation}
Here, $\alpha$ denotes one of the cells belonging to triangle $n$. Note that the value of $\vec{h}(\vec{r})$ does not depend on the choice of $\alpha$ \footnote{This is because from \esref{eq:triangleTransformationTensor} and \seref{eq:triangleMu} follows that if \eref{eq:triangleDefinitionU} holds for one corner of $n$, it also holds for the other two corners.}. The triangle deformation tensor $\itens{u}_{ij}^n$ is defined by
\begin{equation}
  \itens{u}_{ij}^n = \itens{m}_{ji}^n-\itens{\delta}_{ij}\text{.}\label{eq:triangleMu}
\end{equation}
Note the exchanged order of indices at the transformation tensor $\itens{m}_{ji}^n$.
\eref{eq:triangleDefinitionU} defines the displacement field $\vec{h}(\vec{r})$ throughout the entire triangular network such that the displacement gradient is constant on the area of each triangle $n$, taking the value of the triangle deformation tensor: $\partial_i\ivec{h}_j=\itens{u}_{ij}^n$.

Based on this displacement field, the large-scale deformation tensor $\itens{U}_{ij}$ as defined in \eref{eq:UOmegaContinuousArea} can be expressed as the average triangle deformation tensor defined in \eref{eq:triangleMu}:
\begin{equation}
  \itens{U}_{ij} = \left\langle \itens{u}_{ij}\right\rangle\text{.}\label{eq:UOmegaTrianglesArea}
\end{equation}
Here, the brackets denote an area-weighted average:
\begin{equation}
  \big\langle \itens{u}_{ij}\big\rangle = \frac{1}{A}\sum_n{a^n\itens{u}_{ij}^n}\label{eq:definitionAverage}
\end{equation}
with $A$ being the sum of all triangle areas and $a^n$ being the area of triangle $n$. 

Using \eref{eq:UOmegaContinuousBoundary}, the large-scale deformation tensor $\itens{U}_{ij}$ can also be computed from the displacements of cell centers along the margin of the triangular network.
The margin is a chain of triangle sides, and carrying out the boundary integral in \eref{eq:UOmegaContinuousBoundary} for each triangle side, \eref{eq:UOmegaTrianglesArea} can be exactly rewritten as:
\begin{equation}
  \itens{U}_{ij} = \frac{1}{A}\sum_\bondindex{\alpha\beta}{\ivec{h}_j^\bondindex{\alpha\beta}\ivec{\nu}_i^\bondindex{\alpha\beta}\Delta\ell^\bondindex{\alpha\beta}}\text{.}\label{eq:UOmegaTrianglesBoundary}
\end{equation}
Here, $\bondindex{\alpha\beta}$ runs over all triangle sides along the boundary such that cell $\beta$ succeeds cell $\alpha$ in clockwise order, and:
\begin{align}
  \ivec{\nu}^\bondindex{\alpha\beta}_i\Delta\ell^\bondindex{\alpha\beta} &= \itens{\epsilon}_{ik}\ivec{r}^\bondindex{\alpha\beta}_k \\
  \ivec{h}^\bondindex{\alpha\beta}_j &= \frac{1}{2}(\ivec{h}^\alpha_j+\ivec{h}^\beta_j)\text{.}
\end{align}
Thus, the vector $\ivec{\nu}_i^\bondindex{\alpha\beta}$ is the unit vector normal to side $\bondindex{\alpha\beta}$, pointing outside, the scalar $\Delta\ell^\bondindex{\alpha\beta}$ is the length of side $\bondindex{\alpha\beta}$, and the vector $\ivec{h}_j^\bondindex{\alpha\beta}$ is its average displacement.

\section{Triangle shapes and network deformation}
\label{sec:triangles}
We examine the relationship between large-scale deformation and cellular shape changes. To this end, we introduce quantities characterizing the shape of single triangles, and discuss their precise relation to triangle deformation.

\subsection{Elongation of a single triangle}
\label{sec:triangleShape}
Here, we define a symmetric, traceless tensor $\itens{\nem{q}}_{ij}^n$ that characterizes the state of elongation of a triangle $n$. In this and the following section we will omit the subscript $n$ on all triangle-related quantities.

We first introduce a triangle shape tensor $\itens{s}_{ij}$, which maps a virtual equilateral reference triangle to triangle $n$ (\fref{fig:triangleShapeNew}A). More precisely, each side vector $\vec{c}^\bondindex{\alpha\beta}$ of the equilateral reference triangle is mapped to the corresponding side vector $\vec{r}^\bondindex{\alpha\beta}$ of the given triangle $n$:
\begin{equation}
  \ivec{r}^\bondindex{\alpha\beta}_i = \itens{s}_{ij}\ivec{c}^\bondindex{\alpha\beta}_j\text{.}\label{eq:triangleStateTensor}
\end{equation}
The reference triangle has given area $a_0$ and given orientation. Its side vectors $\vec{c}^\bondindex{\alpha\beta}$ are defined in \aref{app:referenceTriangleSides}. Note that \eref{eq:triangleStateTensor} uniquely defines the shape tensor $\itens{s}_{ij}$.

In the special case where the triangle is an equilateral triangle, the elongation tensor is zero, $\itens{\nem{q}}_{ij}=0$. In this case, the shape tensor $\itens{s}_{ij}$ can be expressed as the product of a rotation by a triangle orientation angle $\theta$ and an area scaling (\fref{fig:triangleShapeNew}B):
\begin{equation}
  \tens{s} = \left(\frac{a}{a_0}\right)^{1/2}\tens{\rot}{\left(\theta\right)}\text{.}\label{eq:triangleShapePropertiesQ=0}
\end{equation}
Here, we denote tensors by bold symbols. The tensor $\tens{\rot}{(\theta)}=\exp{(\theta\tens{\epsilon})}$ denotes a counter-clockwise rotation by $\theta$, where the exponential of a tensor is defined by the Taylor series of the exponential function \footnote{Note that \eref{eq:triangleShapePropertiesQ=0} defines the triangle orientation angle $\theta$ modulo $2\pi/3$, because of the different possible associations of the corners of the reference triangle to the corners of triangle $n$. We require the associations between the triangle corners to be made going around both triangles in the same order -- either clockwisely or counter-clockwisely.}.

In the case of a general triangle with nonzero elongation, $\itens{\nem{q}}_{ij}\neq0$, we need an additional anisotropic, area-preserving transformation, i.e.\ a pure shear transformation. This pure shear transformation defines the elongation tensor $\itens{\nem{q}}_{ij}$ (\fref{fig:triangleShapeNew}C):
\begin{equation}
  \tens{s} = \left(\frac{a}{a_0}\right)^{1/2}\exp{(\tens{\nem{q}})}\cdot\tens{\rot}{\left(\theta\right)}\text{.}\label{eq:triangleShapeProperties}
\end{equation}
The dot denotes the tensor product. Note that the exponential of a symmetric, traceless tensor has determinant one and describes a pure shear transformation.
Also note that for given $\itens{s}_{ij}$, \eref{eq:triangleShapeProperties} uniquely defines triangle area $a$, triangle elongation $\itens{\nem{q}}_{ij}$, and the absolute triangle orientation angle $\theta$ (see \aref{app:polarDecomposition}, \citep{Merkel2014b}).

Norm and axis of the elongation tensor
\begin{equation}
  \tens{\nem{q}}=\norm{\itens{\nem{q}}}\begin{pmatrix} \cos{(2\phi)} & \sin{(2\phi)} \\ \sin{(2\phi)} & -\cos{(2\phi)} \end{pmatrix} \label{eq:normOfQ}
\end{equation}
are given by $\norm{\itens{\nem{q}}}=[(\itens{\nem{q}}_{xx})^2+(\itens{\nem{q}}_{xy})^2]^{1/2}=[\trace{(\tens{\nem{q}}^2)}/2]^{1/2}$ and the angle $\phi$ (see \aref{app:interpretationQ}).

Note that the pure shear transformation $\exp{(\tens{\nem{q}})}$ and the rotation $\tens{\rot}{\left(\theta\right)}$ in \eref{eq:triangleShapeProperties} do not commute. Exchanging both in \eref{eq:triangleShapeProperties} leads to a different definition of the elongation angle $\phi\mapsto\phi-\theta$, whereas the elongation norm $\norm{\itens{\nem{q}}}$ and the triangle orientation angle $\theta$ remain unchanged.

\subsection{Triangle deformations corresponding to triangle shape changes}
\label{sec:triangleDeformationAndShape}
\begin{figure}
  \centering
  \includegraphics{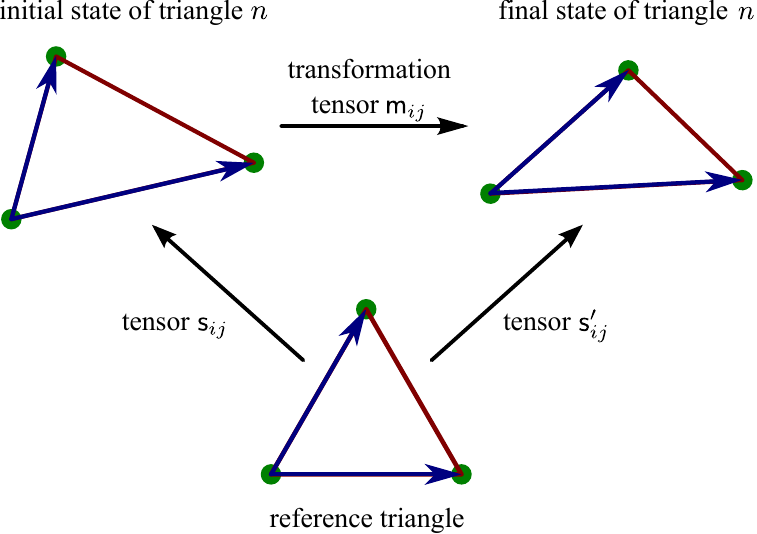}
  \caption{Connection between triangle shape and triangle deformation. A triangle deforms from an initial state to a final state. Deformation, initial state, and final state are characterized by the tensors $\itens{m}_{ij}$, $\itens{s}_{ij}$, and $\itens{s}_{ij}^{\prime}$, respectively.\label{fig:triangleDeformationAndShapeChange}}
\end{figure}
To reveal the precise relationship between triangle deformation and triangle shape, we consider again the deformation of a triangle $n$, which is characterized by the tensor $\itens{m}_{ij}$ (\fref{fig:triangleDeformationAndShapeChange}). We denote the initial and final shape tensors of the triangle by $\itens{s}_{ij}$ and $\itens{s}_{ij}^{\prime}$, respectively. Since both shape tensors are defined with respect to the same reference triangle, the following relation holds:
\begin{equation}
  \itens{s}_{ij}^{\prime} = \itens{m}_{ik}\,\itens{s}_{kj}\text{.}\label{eq:connectionTriangleShapeAndTransformation}
\end{equation}
Based on this equation, the triangle deformation tensor $\itens{u}_{ij}$ can be expressed in terms of triangle shape change.
We define trace $\itens{u}_{kk}$, symmetric, traceless part $\itens{\unem{u}}_{ij}$, and antisymmetric part $\psi$ of the triangle deformation tensor $\itens{u}_{ij}$ as in \eref{eq:decompositionU}:
\begin{equation}
  \itens{u}_{ij} = \frac{1}{2}\itens{u}_{kk}\itens{\delta}_{ij} + \itens{\unem{u}}_{ij} - \psi\itens{\epsilon}_{ij}\text{.}\label{eq:decompositionUTriangle}
\end{equation}
Then, for infinitesimal changes $\delta\itens{\nem{q}}_{ij}$, $\delta a$, $\delta\theta$ of the triangle shape properties $\itens{\nem{q}}_{ij}$, $a$, $\theta$, the following relations hold (see \aref{app:triangleshapeAndDeformationDerivation}):
\begin{eqnarray}
  & \delta\itens{\unem{u}}_{ij} &= \delta\itens{\nem{q}}_{ij} + \delta\itens{\nem{j}}_{ij}\label{eq:triangleShearAndQ}\\
  & \delta\itens{u}_{kk} &= \delta(\ln{a})\label{eq:triangleGAndA} \\
  & \delta\psi &= \delta\theta - \delta\xi\text{.}\label{eq:triangleRotationAndTheta}
\end{eqnarray}
Here, the $\delta$ on the left-hand sides indicate that the respective components $\delta\itens{\unem{u}}_{ij}$, $\delta\itens{u}_{kk}$, and $\delta\psi$ of the deformation tensor are infinitesimal. The following infinitesimal contributions appear on the right-hand sides:
\begin{eqnarray}
  & \delta\itens{\nem{j}}_{ij} &= -2\Big[c\delta\psi+(1-c)\delta\phi\Big]\itens{\epsilon}_{ik}\itens{\nem{q}}_{kj}\label{eq:triangleCorotationalTerm} \\
  & \delta\xi &= \delta\itens{\unem{u}}_{ij}\itens{\epsilon}_{jk}\itens{\nem{q}}_{ki}\frac{\cosh{(2\norm{\itens{\nem{q}}})}-1}{2\norm{\itens{\nem{q}}}\sinh{(2\norm{\itens{\nem{q}}})}}\text{.}\label{eq:triangleShearInducedRotationTerm}
\end{eqnarray}
Here, we have set $c=\tanh{(2\norm{\itens{\nem{q}}})}/2\norm{\itens{\nem{q}}}$, and $\delta\phi$ denotes the change of the elongation axis angle $\phi$.

\subsubsection{Pure shear rate of a single triangle}
To discuss \eref{eq:triangleShearAndQ} relating triangle shear to triangle elongation, we consider that the infinitesimal deformation occurs during an infinitesimal time interval $\delta t$. Then, the triangle pure shear rate $\itens{\unem{v}}_{ij}$ is given by $\itens{\unem{v}}_{ij}\delta t=\delta\itens{\unem{u}}_{ij}$. According to \eref{eq:triangleShearAndQ}, the pure shear rate corresponds exactly to a time derivative of $\itens{\nem{q}}_{ij}$:
\begin{equation}
  \itens{\unem{v}}_{ij}=\frac{\D\itens{\nem{q}}_{ij}}{\D t}\text{.}\label{eq:triangleShearRateAndQ}
\end{equation}
This generalized corotational time derivative is defined by $(\D\itens{\nem{q}}_{ij}/\D t)\delta t = \delta\itens{\nem{q}}_{ij} + \delta\itens{\nem{j}}_{ij}$, which can be rewritten as
\begin{equation}
  \frac{\D\itens{\nem{q}}_{ij}}{\D t} = \frac{\d\itens{\nem{q}}_{ij}}{\d t} -2\left(c\omega+(1-c)\frac{\d\phi}{\d t}\right) \itens{\epsilon}_{ik}\itens{\nem{q}}_{kj}\text{.}\label{eq:triangleMaterialDerivativeQ}
\end{equation}
Here, the operator $\d/\d t$ denotes the total time derivative of a quantity and $\omega$ is the triangle vorticity with $\omega\delta t=\delta\psi$.
In the limit $\norm{\itens{\nem{q}}}\ll 1$ for which $c\simeq 1$, the generalized corotational derivative becomes the conventional Jaumann derivative \cite{Bird1987v2}:
\begin{equation}
  \frac{\D\itens{\nem{q}}_{ij}}{\D t} \simeq \frac{\d\itens{\nem{q}}_{ij}}{\d t} +\itens{\omega}_{ik}\itens{\nem{q}}_{kj}-\itens{\nem{q}}_{ik}\itens{\omega}_{kj}\text{,}\label{eq:triangleMaterialDerivativeQc=1}
\end{equation}
where we introduced $\itens{\omega}_{ij}=-\omega\epsilon_{ij}=(\itens{v}_{ij}-\itens{v}_{ji})/2$. The general case of finite $\norm{\itens{\nem{q}}}$ with $c\neq1$ is discussed in more detail in \aref{app:corotationalDerivative}.

\subsubsection{Isotropic expansion rate and vorticity of a single triangle}
According to \eref{eq:triangleGAndA}, the isotropic triangle expansion rate $\itens{v}_{kk}$ with $\itens{v}_{kk}\delta t=\delta\itens{u}_{kk}$ can be written as:
\begin{equation}
  \itens{v}_{kk} = \frac{1}{a}\,\frac{\d a}{\d t}\text{.}
\end{equation}
The isotropic triangle expansion rate thus corresponds to the relative change rate of the triangle area $a$.

Finally, \eref{eq:triangleRotationAndTheta} states that triangle vorticity $\omega$ can be written as
\begin{equation}
  \omega = \frac{\d\theta}{\d t} - \itens{\unem{v}}_{ij}\itens{\epsilon}_{jk}\itens{\nem{q}}_{ki}\frac{\cosh{(2\norm{\itens{\nem{q}}})}-1}{2\norm{\itens{\nem{q}}}\sinh{(2\norm{\itens{\nem{q}}})}}\text{.}\label{eq:triangleVorticity}
\end{equation}
Hence, the triangle orientation angle $\theta$ may not only change due to a vorticity $\omega$ in the flow field, but also due to local pure shear. This shear-induced triangle rotation appears whenever there is a component of the shear rate tensor $\itens{\unem{v}}_{ij}$ that is neither parallel nor perpendicular to the triangle elongation axis. We discuss this effect of shear-induced rotation in more detail in \aref{app:shearInducedRotation}.

\subsection{Large-scale deformation of a triangular network}
\label{sec:largeScaleDeformation}
To understand how triangle shape properties connect to large-scale deformation of a triangle network, we coarse-grain \esref{eq:triangleShearAndQ}-\seref{eq:triangleRotationAndTheta}.
We focus on the case where the shape properties $\itens{\nem{q}}_{ij}^n$, $a^n$, $\theta^n$ of all involved triangles $n$ change only infinitesimally.
The large-scale deformation tensor of the triangular network can be computed using \eref{eq:UOmegaTrianglesArea}: $\delta\itens{U}_{ij} = \left\langle\delta\itens{u}_{ij}\right\rangle$. Consequently, one obtains large-scale pure shear as $\delta\itens{\unem{U}}_{ij} = \left\langle\delta\itens{\unem{u}}_{ij}\right\rangle$, large-scale isotropic expansion as $\delta\itens{U}_{kk} = \left\langle\delta\itens{u}_{kk}\right\rangle$, and large-scale rotation as $\delta\Psi = \left\langle\delta\psi\right\rangle$. We now express large-scale pure shear and isotropic expansion in terms of triangle shape changes. We discuss large-scale rotation in \aref{app:rotationTriangleNetwork}.

\subsubsection{Pure shear deformation on large scales}
\label{sec:largeScalePureShear}
To discuss large-scale pure shear deformation, we first introduce an average triangle elongation tensor:
\begin{equation}
  \itens{\nem{Q}}_{ij} = \left\langle\itens{\nem{q}}_{ij}\right\rangle\text{.}
\end{equation}
The average is computed using an area weighting as in \eref{eq:definitionAverage}.

The large-scale pure shear tensor $\delta\itens{\unem{U}}_{ij}$ can be related to the change of the average triangle elongation $\delta\itens{\nem{Q}}_{ij}$ by averaging \eref{eq:triangleShearAndQ} over all triangles in the triangulation (see \aref{app:largeScaleShear}):
\begin{equation}
  \delta\itens{\unem{U}}_{ij} = \delta\itens{\nem{Q}}_{ij} + \delta\itens{\nem{J}}_{ij} + \delta\itens{\nem{K}}_{ij}\text{.}\label{eq:decompositionShearWithoutTopologicalTransitions}
\end{equation}
Here, we introduced the mean-field corotational term
\begin{equation}
  \delta\itens{\nem{J}}_{ij} = -2\Big[C\delta\Psi+(1-C)\delta\Phi\Big]\itens{\epsilon}_{ik}\itens{\nem{Q}}_{kj}\text{,}\label{eq:corotationalTerm}
\end{equation}
where $C=\tanh{(2\norm{\itens{\nem{Q}}})}/2\norm{\itens{\nem{Q}}}$, and $\norm{\itens{\nem{Q}}}$ and $\Phi$ denote norm and angle of the average elongation tensor $\itens{\nem{Q}}_{ij}$, respectively. Moreover, the contribution $\delta\itens{\nem{K}}_{ij}$ newly appears due to the averaging. It is the sum of two correlations:
\begin{equation}
  \delta\itens{\nem{K}}_{ij} = - \Big(\big\langle\delta\itens{u}_{kk}\itens{\nem{q}}_{ij}\big\rangle-  \delta\itens{U}_{kk}\itens{\nem{Q}}_{ij}\Big) + \Big(\langle\delta\itens{\nem{j}}_{ij}\rangle - \delta\itens{\nem{J}}_{ij}\Big)\text{.}\label{eq:definitionCorrelationTerm}
\end{equation}
We call the first term \textit{growth correlation} and the second term \textit{rotational correlation}.

\begin{figure}
  \centering
  \includegraphics{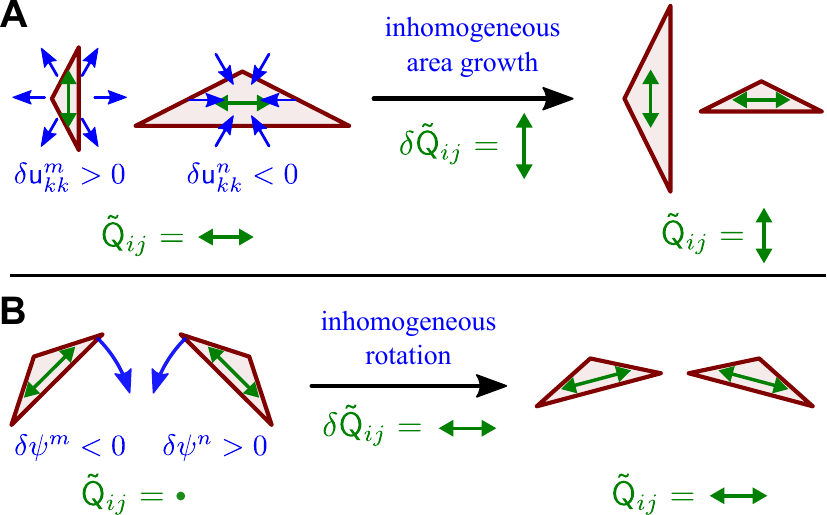}
  \caption{Correlation contributions to pure shear.
  (A) Inhomogeneous isotropic expansion that is correlated with elongation creates a change in the average elongation $\itens{\nem{Q}}_{ij}$, which is due to the area weighting in the definition of $\itens{\nem{Q}}_{ij}$. This contribution to the time derivative of $\itens{\nem{Q}}_{ij}$ is compensated for by the growth correlation term in $\delta\itens{\nem{K}}_{ij}$. 
  (B) Inhomogeneous rotation that is correlated with elongation creates a change in the average elongation $\itens{\nem{Q}}_{ij}$. This contribution to the time derivative of $\itens{\nem{Q}}_{ij}$ is compensated for by the rotational correlation term in $\delta\itens{\nem{K}}_{ij}$.\label{fig:correlations}}
\end{figure}
Growth correlation is created by spatial fluctuations in isotropic triangle expansion $\delta\itens{u}_{kk}^n$. \Fref{fig:correlations}A illustrates this effect for a deformation where no large-scale pure shear appears $\delta\itens{\unem{U}}_{ij}=0$. 
Two triangles with different but constant triangle elongation tensors $\itens{\nem{q}}_{ij}^n$ deform: One triangle expands isotropically and the other triangle shrinks isotropically. Because of the area-weighting in the averaging, the average elongation tensor $\itens{\nem{Q}}_{ij}$ thus changes during this deformation. Therefore, although $\delta\itens{\unem{U}}_{ij}=0$ in \eref{eq:decompositionShearWithoutTopologicalTransitions}, the average elongation changes by $\delta\itens{\nem{Q}}_{ij}\neq 0$. This change in average elongation is exactly compensated for by the growth correlation term.

Rotational correlation can be created by spatial fluctuations of triangle rotation $\delta\psi^n$. We illustrate this in \fref{fig:correlations}B, where the large-scale pure shear rate is again zero $\delta\itens{\unem{U}}_{ij}=0$. We consider two triangles with the same area but different elongation tensors $\itens{\nem{q}}_{ij}^n$. Both triangles do not deform, but rotate in opposing directions by the same absolute angle $\delta\psi^n$. The large-scale corotational term is zero $\delta\itens{\nem{J}}_{ij}=0$, because there is no overall rotation $\delta\Psi=0$. However, the corotational term for each individual triangle $\delta\itens{\nem{j}}_{ij}^n$ is nonzero allowing for a change of triangle elongation in the absence of triangle shear. After all, the average elongation tensor $\itens{\nem{Q}}_{ij}$ increases along the horizontal, because each individual triangle elongation tensor does. 
This change in average elongation is compensated for by the rotational correlation term.

To obtain the large-scale pure shear rate $\itens{\unem{V}}_{ij}$ defined by $\itens{\unem{V}}_{ij}\delta t=\delta\itens{\unem{U}}_{ij}$, we rewrite \eref{eq:decompositionShearWithoutTopologicalTransitions}:
\begin{equation}
  \itens{\unem{V}}_{ij} = \frac{\D\itens{\nem{Q}}_{ij}}{\D t} + \itens{\nem{D}}_{ij}\text{.} \label{eq:decompositionShearRateWithoutTopologicalTransitions}
\end{equation}
Here, $\D\itens{\nem{Q}}_{ij}/\D t$ denotes a corotational time derivative that is defined by $(\D\itens{\nem{Q}}_{ij}/\D t)\delta t = \delta\itens{\nem{Q}}_{ij} + \delta\itens{\nem{J}}_{ij}$, which can be rewritten as
\begin{equation}
  \frac{\D\itens{\nem{Q}}_{ij}}{\D t} = \frac{\d\itens{\nem{Q}}_{ij}}{\d t} -2\left(C\Omega+(1-C)\frac{\d\Phi}{\d t}\right) \itens{\epsilon}_{ik}\itens{\nem{Q}}_{kj}\text{.}\label{eq:triangleMaterialDerivativeAverageQ}
\end{equation}
Here, $C=\tanh{(2\norm{\itens{\nem{Q}}})}/2\norm{\itens{\nem{Q}}}$ as defined below \eref{eq:corotationalTerm} and $\Omega$ is the average vorticity with $\omega\delta t=\delta\Psi$.
The term $\itens{\nem{D}}_{ij}$ in \eref{eq:decompositionShearRateWithoutTopologicalTransitions} contains the correlation terms with $\itens{\nem{D}}_{ij}\delta t=\delta\itens{\nem{K}}_{ij}$.

\eref{eq:decompositionShearRateWithoutTopologicalTransitions} is an important result for the case without topological transitions. It states that the large-scale deformation of a triangular network can be computed from the change of the average triangle elongation, the correlation between triangle elongation and triangle area growth, and the correlation between triangle elongation and triangle rotation.

The correlations account for the fact that taking the corotational derivative does not commute with averaging:
\begin{equation}
  \itens{\nem{D}}_{ij} = \bigg\langle\frac{\D\itens{\nem{q}}_{ij}}{\D t}\bigg\rangle - \frac{\D\itens{\nem{Q}}_{ij}}{\D t}\text{.}\label{eq:correlationTermIsCommutator}
\end{equation}
In particular, as illustrated in \fref{fig:correlations}B, the rotational correlation arises by coarse-graining of the corotational term. Similarly, the growth correlation can be regarded as arising from the coarse-graining of a convective term (see \aref{app:correlationConvection}).

\subsubsection{Elongation and shear of a single cell}
\begin{figure}
  \centering
  \includegraphics{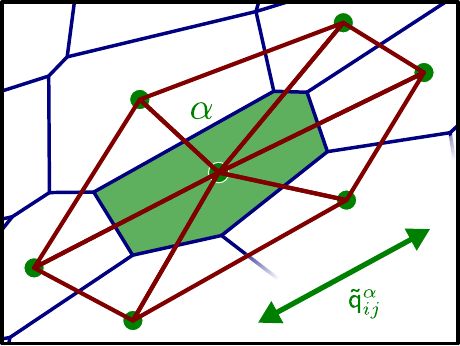}
  \caption{The elongation $\itens{\nem{q}}^\alpha_{ij}$ of a cell $\alpha$ (green) is defined by the average elongation of the triangles belonging to $\alpha$ (red). The triangles belonging to $\alpha$ are those that have one of their corners defined by the center of $\alpha$. \label{fig:cellElongation}}
\end{figure}
To more explicitly relate the above discussion to cell shape and deformation, we define a cell elongation tensor $\itens{\nem{q}}_{ij}^\alpha$ for a given cell $\alpha$ as follows. We select all triangles $n$ that have one of their corners defined by the center of $\alpha$, and then average their elongation tensors (\fref{fig:cellElongation}):
\begin{equation}
  \itens{\nem{q}}_{ij}^\alpha = \left\langle\itens{\nem{q}}_{ij}\right\rangle\text{.}
\end{equation}
The average is again area-weighted as defined in \eref{eq:definitionAverage}. 
Then, a cellular pure shear rate can be defined analogously: $\itens{\unem{v}}_{ij}^\alpha = \left\langle\itens{\unem{v}}_{ij}\right\rangle$. This cellular pure shear rate can also be expressed by changes of $\itens{\nem{q}}_{ij}^\alpha$ using \eref{eq:decompositionShearRateWithoutTopologicalTransitions}. Moreover, the large-scale elongation $\itens{\nem{Q}}_{ij}$ and the large-scale pure shear rate $\itens{\unem{V}}_{ij}$ can be obtained by suitably averaging the single-cell quantities $\itens{\nem{q}}_{ij}^\alpha$ and $\itens{\unem{v}}_{ij}^\alpha$ \footnote{For such an average, the cellular quantities $\itens{\nem{q}}_{ij}^\alpha$ and $\itens{\unem{v}}_{ij}^\alpha$ have to be weighted by the summed area $a^\alpha_\triangle=\sum_na^n$ of all triangles $n$ belonging to the respective cell $\alpha$. Up to boundary terms these averages then respectively correspond to the large-scale quantities $\itens{\nem{Q}}_{ij}$ and $\itens{\unem{V}}_{ij}$.}.

\subsubsection{Isotropic expansion on large scales}
\label{sec:largeScaleIsotropicExpansion}
Finally, we discuss large-scale isotropic expansion $\delta\itens{U}_{kk}$ of a triangle network. We relate it to changes of the average triangle area $\bar{a}=A/N$, where $A$ is the total area of the network and $N$ is the number of triangles in the network.

To relate large-scale isotropic expansion $\delta\itens{U}_{kk}$ to changes of the average triangle area $\bar{a}$, we average \eref{eq:triangleGAndA}:
\begin{equation}
  \delta\itens{U}_{kk} = \delta(\ln{\bar{a}})\text{.}\label{eq:decompositionExpansionWithoutTopologicalTransitions}
\end{equation}
Accordingly, the large-scale isotropic expansion rate $\itens{V}_{kk}$ with $\itens{V}_{kk}\delta t=\delta\itens{U}_{kk}$ can be expressed as
\begin{equation}
  \itens{V}_{kk} = \frac{1}{\bar{a}}\,\frac{\d \bar{a}}{\d t}\text{.}\label{eq:decompositionExpansionRateWithoutTopologicalTransitions}
\end{equation}
Hence, large-scale isotropic expansion corresponds to the relative change of the average triangle area $\bar{a}$.

\section{Contributions of topological transitions to network deformation}
\label{sec:ttcontributionsToDeformation}
So far, we have considered deformations of a triangular network during which no topological transitions occur. 
Now, we discuss the contributions of topological transitions to large-scale deformations \footnote{More precisely, here and in the following, we consider topological transitions occurring in bulk. For a discussion of topological transitions occurring at the margin of the polygonal network, i.e.\ topological transitions altering the sequence of cell centers that forms the margin of the triangulation, see \cite{Merkel2014b}.}.

\begin{figure}
  \centering
  \includegraphics{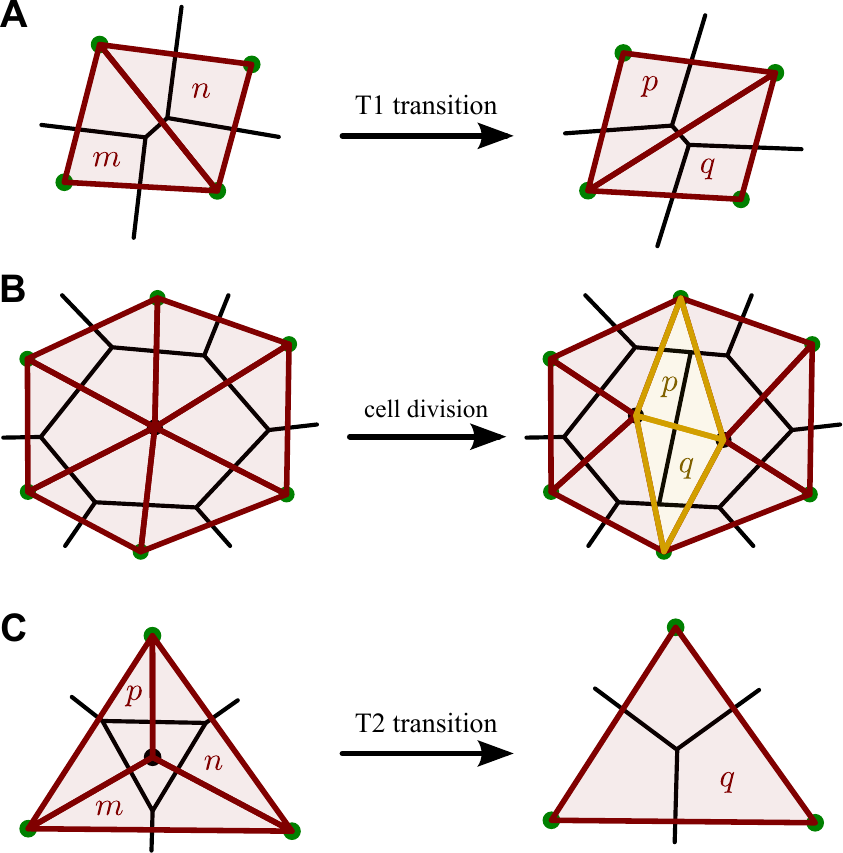}
  \caption{Effects of a single topological transition on the triangulation. (A) A T1 transition removes two triangles ($m$ and $n$) and creates two new ones ($p$ and $q$). (B) A cell division creates two triangles ($p$ and $q$, yellow). All other triangles shown (red) change their shape instantaneously. (C) A T2 transition removes three triangles ($m$, $n$, $p$) and creates a new one ($q$).\label{fig:topologicalTransitions}}
\end{figure}
There are two main features of topological transitions that motivate the following discussion. First, topological transitions occur instantaneously at precise time points $t_k$ and correspondingly, there is no displacement of cell centers upon topological transitions.

Second, topological transitions create and remove triangles from the triangulation. For instance for the typical case of three-fold vertices, a T1 transition removes two triangles and then adds two new triangles (\fref{fig:topologicalTransitions}A), a cell division just adds two triangles (\fref{fig:topologicalTransitions}B), and a T2 transitions removes three triangles and adds one new triangle (\fref{fig:topologicalTransitions}C).

To define the large-scale deformation tensor across a given topological transition, an average over triangle deformations as in \eref{eq:UOmegaTrianglesArea} can no longer be used because the triangle deformation tensor $\itens{u}_{ij}^n$ is ill-defined for disappearing and appearing triangles. We thus define the large-scale deformation depending on cell center displacements along the margin of the triangular network using \eref{eq:UOmegaTrianglesBoundary}. We denote such a large-scale deformation tensor across a topological transition by $\Delta\itens{U}_{ij}$.
Because there are no cell center displacements upon a topological transition, the large-scale deformation tensor vanishes $\Delta\itens{U}_{ij}=0$, and so does large-scale isotropic expansion $\Delta\itens{U}_{kk}=0$ and large-scale pure shear $\Delta\itens{\unem{U}}_{ij}=0$.
However, even though there is no actual network deformation \textit{upon} a topological transition, we will define the deformation contribution \textit{by} a topological transition in the following.

\subsection{Contribution of a single topological transition to pure shear}
\label{sec:ttcontributionsToShear}
\begin{figure}
  \centering
  \includegraphics{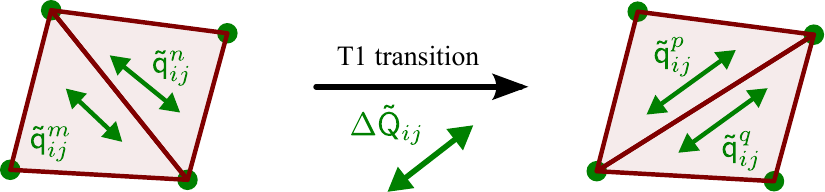}
  \caption{A T1 transition induces an instantaneous change of the average triangle elongation. The average triangle elongation before and after the T1 transition only depends on the position of the four involved cell centers (green dots).\label{fig:T1transitionQ}}
\end{figure}
To discuss the pure shear contribution by a topological transition, we focus on a single T1 transition occurring at time $t_k$. Pure shear contributions by cell divisions or T2 transitions can be discussed analogously.

\begin{figure*}
  \centering
  \includegraphics{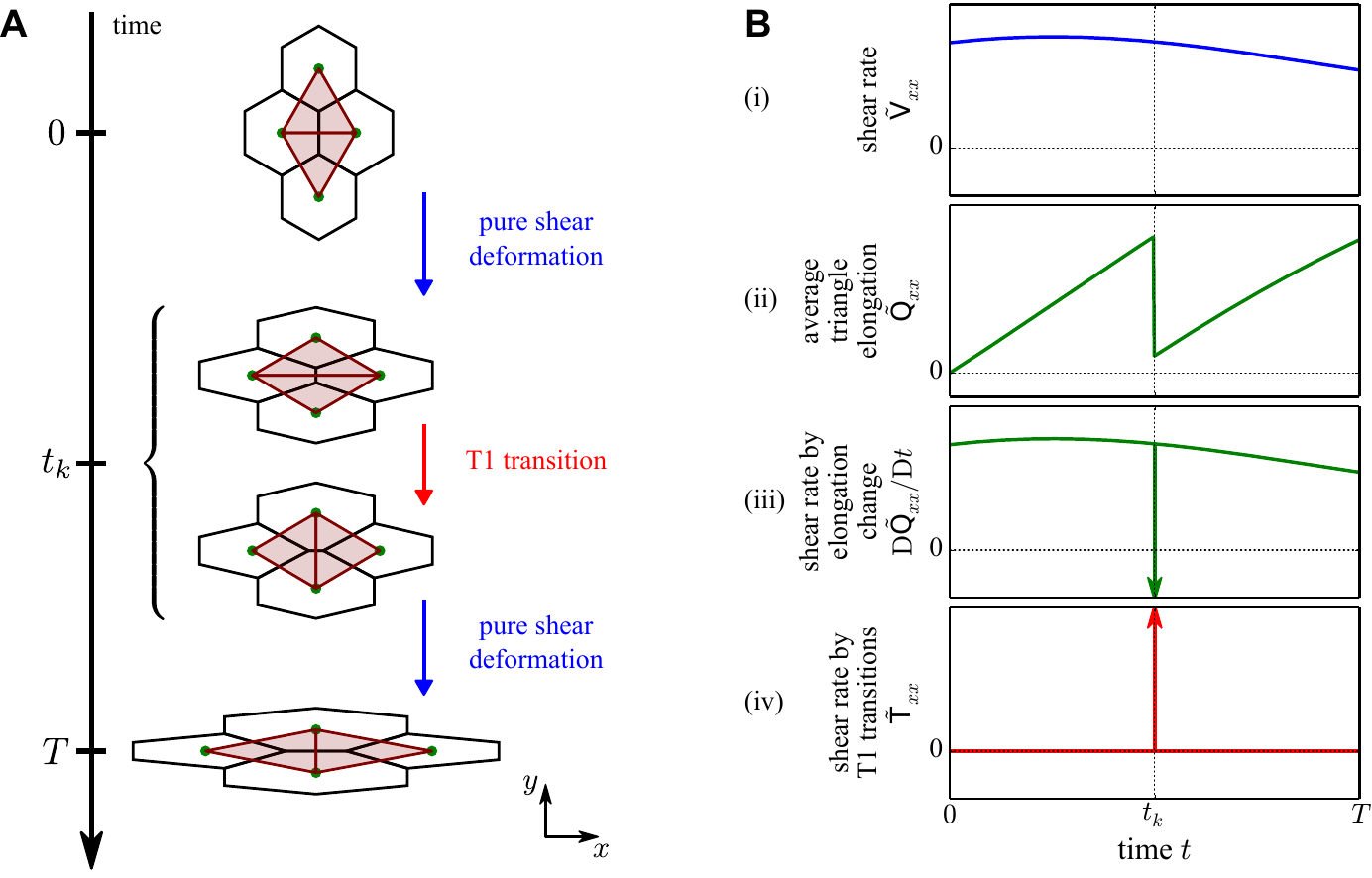}
  \caption{Illustration of the shear rate contributions by average triangle elongation change and T1 transitions. (A) Between the time points $0$ and $T$, a triangular network is continously sheared along the horizontal axis. At time point $t_k$, a T1 transition occurs, which instantaneously changes the average triangle elongation. (B) For the process shown in panel A, schematic time-dependent plots of shear rate (blue, i), the average triangle elongation (green, ii), its derivative (green, iii), and the shear rate by T1 transitions (red, iv). For each tensor, the respective $xx$ component, i.e.\ the horizontal component, is plotted. The arrows in (iii) and (iv) indicate Dirac $\delta$ peaks. Their magnitude corresponds to the step in $\itens{\nem{Q}}_{xx}$ at $t_k$. \label{fig:shearRateContributionsIllustration}}
\end{figure*}
Because of the triangulation change during a T1 transition, the average triangle elongation $\itens{\nem{Q}}_{ij}$ changes instantaneously by a finite amount $\Delta\itens{\nem{Q}}_{ij}$ (\fref{fig:T1transitionQ}).
To account for the shear contribution by the T1 transition, we introduce an additional term $\Delta\itens{\nem{X}}_{ij}$ into the shear balance \eref{eq:decompositionShearWithoutTopologicalTransitions}:
\begin{equation}
  \Delta\itens{\unem{U}}_{ij} = \Delta\itens{\nem{Q}}_{ij} + \Delta\itens{\nem{X}}_{ij}\text{.}\label{eq:T1shear}
\end{equation}
Here, we have set corotational and correlation terms during the T1 transition to zero \footnote{Note that this is a convention and that different choices are possible as well (see \aref{app:T1shear}).}.
Because $\Delta\itens{\unem{U}}_{ij}=0$, we obtain from \eref{eq:T1shear} that $\Delta\itens{\nem{X}}_{ij} = -\Delta\itens{\nem{Q}}_{ij}$.
Thus, the shear contribution $\Delta\itens{\nem{X}}_{ij}$ due to the T1 transition compensates for the finite discontinuity in $\itens{\nem{Q}}_{ij}$, which occurs due to the removal and addition of triangles.

Dividing by a time interval $\Delta t$ and in the limit $\Delta t\rightarrow 0$, we can transform \eref{eq:T1shear} into an equation for the shear rate:
\begin{equation}
  \itens{\unem{V}}_{ij} = \frac{\D\itens{\nem{Q}}_{ij}}{\D t} + \itens{\nem{D}}_{ij} + \itens{\nem{T}}_{ij}\text{,}\label{eq:T1shearRate}
\end{equation}
where $\itens{\nem{T}}_{ij}=\Delta\itens{\nem{X}}_{ij}\delta(t-t_k)$ and $\delta$ denotes the Dirac delta function. Hence, a T1 transition induces a discontinuity in the average triangle elongation $\itens{\nem{Q}}_{ij}$, causing a delta peak in $\D\itens{\nem{Q}}_{ij}/\D t$. This delta peak is exactly compensated for by $\Delta\itens{\nem{X}}_{ij}\delta(t-t_k)$, such that the large-scale shear rate $\itens{\unem{V}}_{ij}$ contains no delta peaks.

As an example, \fref{fig:shearRateContributionsIllustration}A illustrates a process during which a network consisting of two triangles (red) is being deformed between the times $0$ and $T$. These triangles undergo a pure shear deformation along the $x$ axis without any rotations or inhomogeneities. In the absence of any topological transition, the shear rate along the $x$ axis, $\itens{\unem{V}}_{xx}$, corresponds to the derivative of the average triangle elongation, $\d\itens{\nem{Q}}_{xx}/\d t$ (\fref{fig:shearRateContributionsIllustration}B(i-iii)). However, at a time point $t_k$, a T1 transition occurs and the average elongation along the $x$ axis changes instantaneously by $\Delta\itens{\nem{Q}}_{xx}$. Thus, there is a Dirac $\delta$ peak in $\d\itens{\nem{Q}}_{xx}/\d t$, which is compensated by the T1 shear rate $\itens{\nem{T}}_{xx}=-\Delta\itens{\nem{Q}}_{xx}\delta(t-t_k)$ (\fref{fig:shearRateContributionsIllustration}B(iv)) such that \eref{eq:T1shearRate} holds exactly.

For the special case where the four cell centers involved in the T1 transition (green dots in \fref{fig:T1transitionQ}) form a square, the magnitude of $\Delta\itens{\nem{X}}_{ij}$ evaluates exactly to $\norm{\Delta\itens{\nem{X}}} = (A_\square\ln{3})/(2A)$, where $A_\square$ is the area of the square and $A$ is the total area of the triangle network (see \aref{app:T1shear:rhombus}). The axis of $\Delta\itens{\nem{X}}_{ij}$ is along one of the diagonals of the square. Both remain true for the more general case of a rhombus, i.e.\ a quadrilateral whose four sides have equal lengths.

\subsection{Contribution of a single topological transition to isotropic expansion}
\label{sec:isotropicExpansionSingleTt}
To define the isotropic expansion by a topological transition, we employ a similar argument as for the pure shear component.
For instance, to account for the isotropic expansion by a single cell division occurring at time $t_k$, we introduce a term $\Delta d$ into \eref{eq:decompositionExpansionWithoutTopologicalTransitions} (cell extrusions can be treated analogously):
\begin{equation}
  \Delta\itens{U}_{kk} = \Delta(\ln{\bar{a}}) + \Delta d\text{.}\label{eq:isotropicExpansionWithCD}
\end{equation}
Here, $\Delta(\ln{\bar{a}})$ denotes the change of $\ln{\bar{a}}$ across the cell division. Since there is no isotropic expansion upon the cell division $\Delta\itens{U}_{kk}=0$, we thus have $\Delta d=-\Delta(\ln{\bar{a}})$. Because the total area $A$ of the triangulation remains constant during the cell division, the isotropic expansion by a cell division amounts to $\Delta d=\ln{(1+2/N)}$ with $N$ being the number of triangles in the network before the division.


Dividing by a time interval $\Delta t$ and in the limit $\Delta t\rightarrow 0$, \eref{eq:isotropicExpansionWithCD} transforms into:
\begin{equation}
  \itens{V}_{kk} = \frac{\d(\ln{\bar{a}})}{\d t} + k_d
\end{equation}
with $k_d=\ln{(1+2/N)}\,\delta(t-t_k)$. Hence, as for the pure shear component, the contributions of individual topological transitions to the isotropic expansion component can be accounted for by delta peaks. 

Note that in order to avoid isotropic expansion contributions by T1 transitions, care has to be taken when counting the number of triangles $N$ for the special case of $M$-fold vertices with $M>3$. In \aref{app:triangulation}, we explain how we define $N$ in this case.

\section{Cellular contributions to the large-scale deformation rate}
\label{sec:allTogether}
We wrap up the previous sections providing equations that express large-scale pure shear and isotropic expansion as sums of all cellular contributions.
To this end, we consider the deformation of a triangle network with an arbitrary number of topological transitions. 
Large-scale rotation is discussed in \aref{app:rotationTriangleNetwork}.

\subsection{Pure shear rate}
We decompose the instantaneous large-scale shear rate $\itens{\unem{V}}_{ij}$ into the following cellular contributions:
\begin{equation}
  \itens{\unem{V}}_{ij} = \frac{\D\itens{\nem{Q}}_{ij}}{\D t} + \itens{\nem{T}}_{ij} + \itens{\nem{C}}_{ij} + \itens{\nem{E}}_{ij} + \itens{\nem{D}}_{ij}\text{.}\label{eq:decompositionPureShear}
\end{equation}
The first term on the right-hand side denotes the corotational time derivative of $\itens{\nem{Q}}_{ij}$ defined by \eref{eq:triangleMaterialDerivativeAverageQ}. Note that some care has to be taken when evaluating the corotational term in the presence of topological transitions (see \aref{app:T1shear}).
The shear rate contributions by T1 transitions $\itens{\nem{T}}_{ij}$, cell divisions $\itens{\nem{C}}_{ij}$, and T2 transitions $\itens{\nem{E}}_{ij}$ to the large-scale shear rate are respectively defined by
\begin{eqnarray}
 &  \itens{\nem{T}}_{ij} &= -\sum_{k\in\mathrm{T1}}{ \Delta\itens{\nem{Q}}^k_{ij} \delta(t-t_k) }\label{eq:shearByT1Transitions} \\
 &  \itens{\nem{C}}_{ij} &= -\sum_{k\in\mathrm{CD}}{ \Delta\itens{\nem{Q}}^k_{ij} \delta(t-t_k) }\label{eq:shearByCellDivisions} \\
 &  \itens{\nem{E}}_{ij} &= -\sum_{k\in\mathrm{T2}}{ \Delta\itens{\nem{Q}}^k_{ij} \delta(t-t_k) }\label{eq:shearByT2Transitions}\text{.}
\end{eqnarray}
Here, the sums run over all topological transitions $k$ of the respective kind, $t_k$ denotes the time point of the respective transition, and $\Delta\itens{\nem{Q}}^k_{ij}$ denotes the instantaneous change in $\itens{\nem{Q}}_{ij}$ induced by the transition.
Finally, $\itens{\nem{D}}_{ij}$ denotes the shear rate by the correlation effects as introduced in \sref{sec:largeScalePureShear}.

\begin{figure*}
  \centering
  \includegraphics{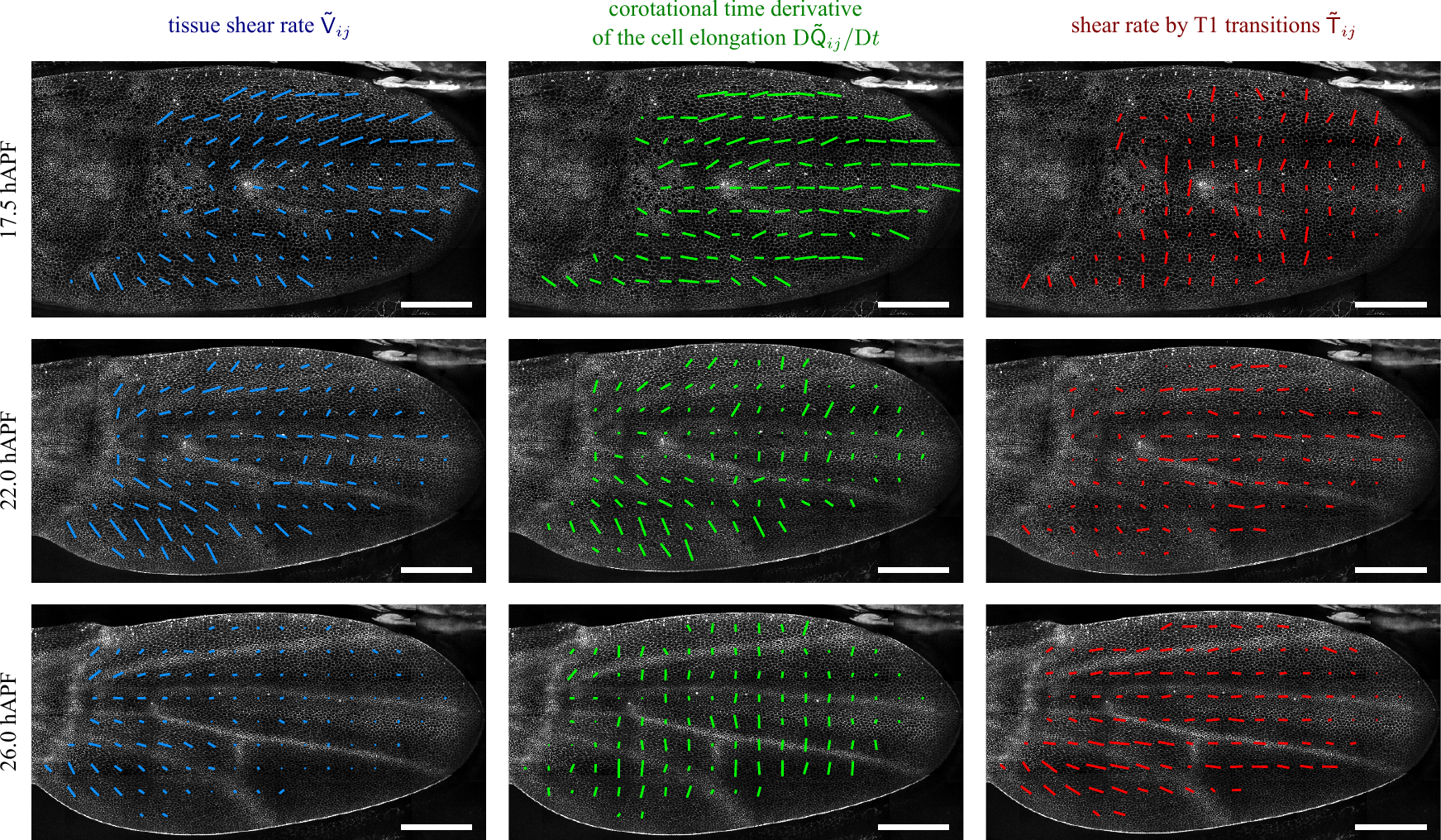}
  \caption{Patterns of tissue shear and contributions to shear in the pupal wing of the fruit fly at different times in hours after puparium formation (hAPF).
  	Local rate of pure shear (blue), corotational time derivative of the cell elongation (center), and shear rate by T1 transitions (right). The bars indicate the axis and norm of the tensors. Shown are averages over squares with size $(33\,\micro\meter)^2$ and over time intervals of about $2\,\hour$. The scale bars correspond to $100\,\micro\metre$.\label{fig:flyWing_patterns}}
\end{figure*}
\begin{figure*}
  \centering
  \includegraphics{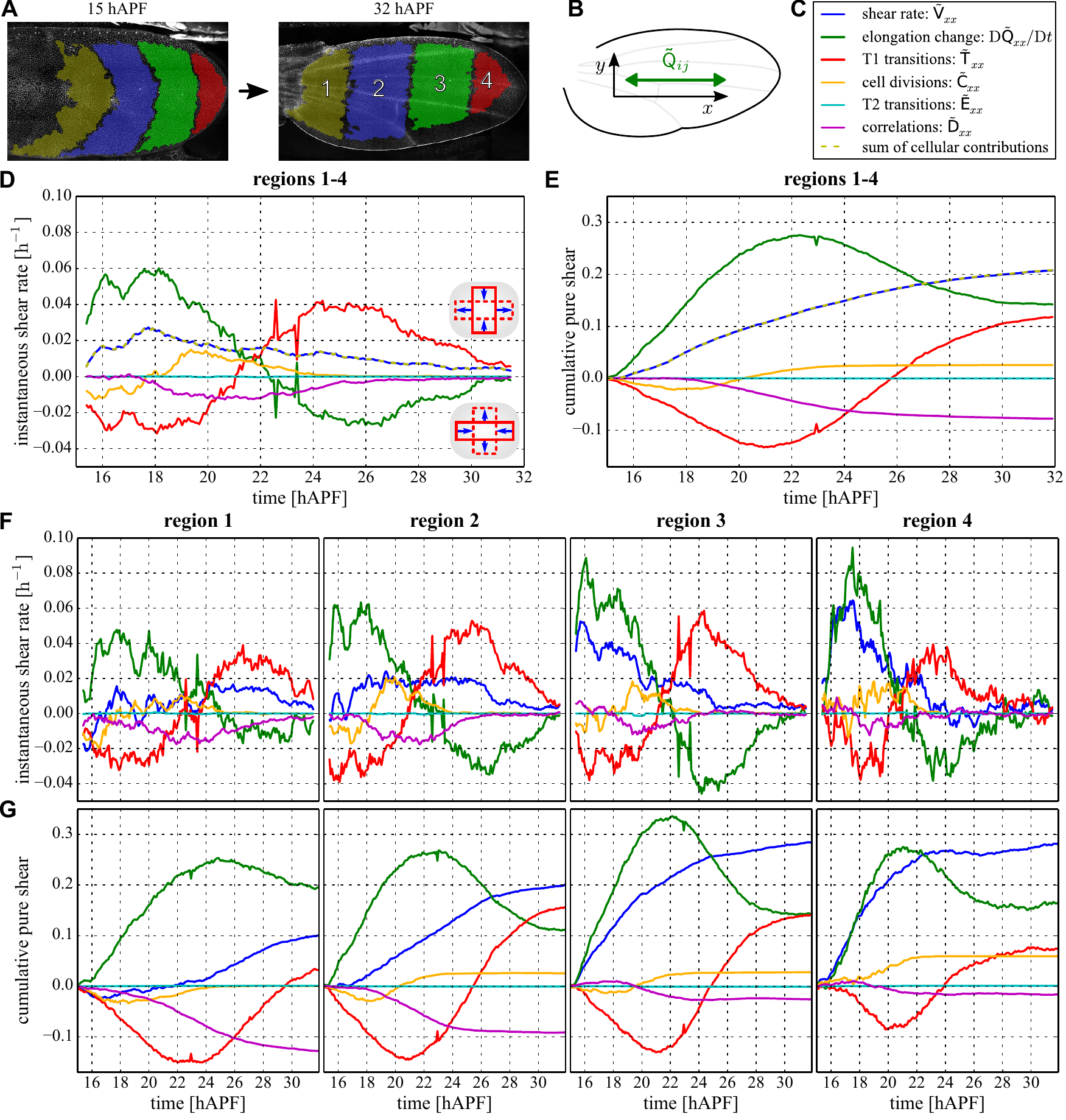}
  \caption{Contributions to tissue shear as a function of time during pupal development of the fly wing. Shown is the data for one wing.
  (A) The fly wing undergoes complex tissue remodeling, which we recorded between 15 and 32 hAPF. The colored areas mark regions of tissue in which all cells were tracked during this time interval.
  (B) Schematic representation of the coordinate system used to describe tissue deformations. The $x$ axis points towards the tip of the wing and is aligned parallel to the axis of cell elongation averaged within the interval between 24 and 32 hAPF and over all four regions. The average cell elongation computed for a single region deviates at most by 5 degrees from this $x$ axis. 
  (C) Legend specifying different contributions to tissue shear.
  (D) Cellular contributions to shear and total shear rate averaged over regions 1-4 in panel A as a function of time. Plotted are the projections of the tensors on the $x$ axis, for example the component $\itens{\unem{V}}_{xx}$ of the tissue shear rate.
  (E) Cumulative tissue shear and cellular contributions, projected on the $x$ axis.
  (F,G) Same plots as in D and E, but for the subregions 1 to 4 indicated in panel A. 
  In D-G, data was averaged over 10 subsequent inter-frame intervals.
  \label{fig:flyWing}}
\end{figure*}
\begin{figure*}
  \centering
  \includegraphics{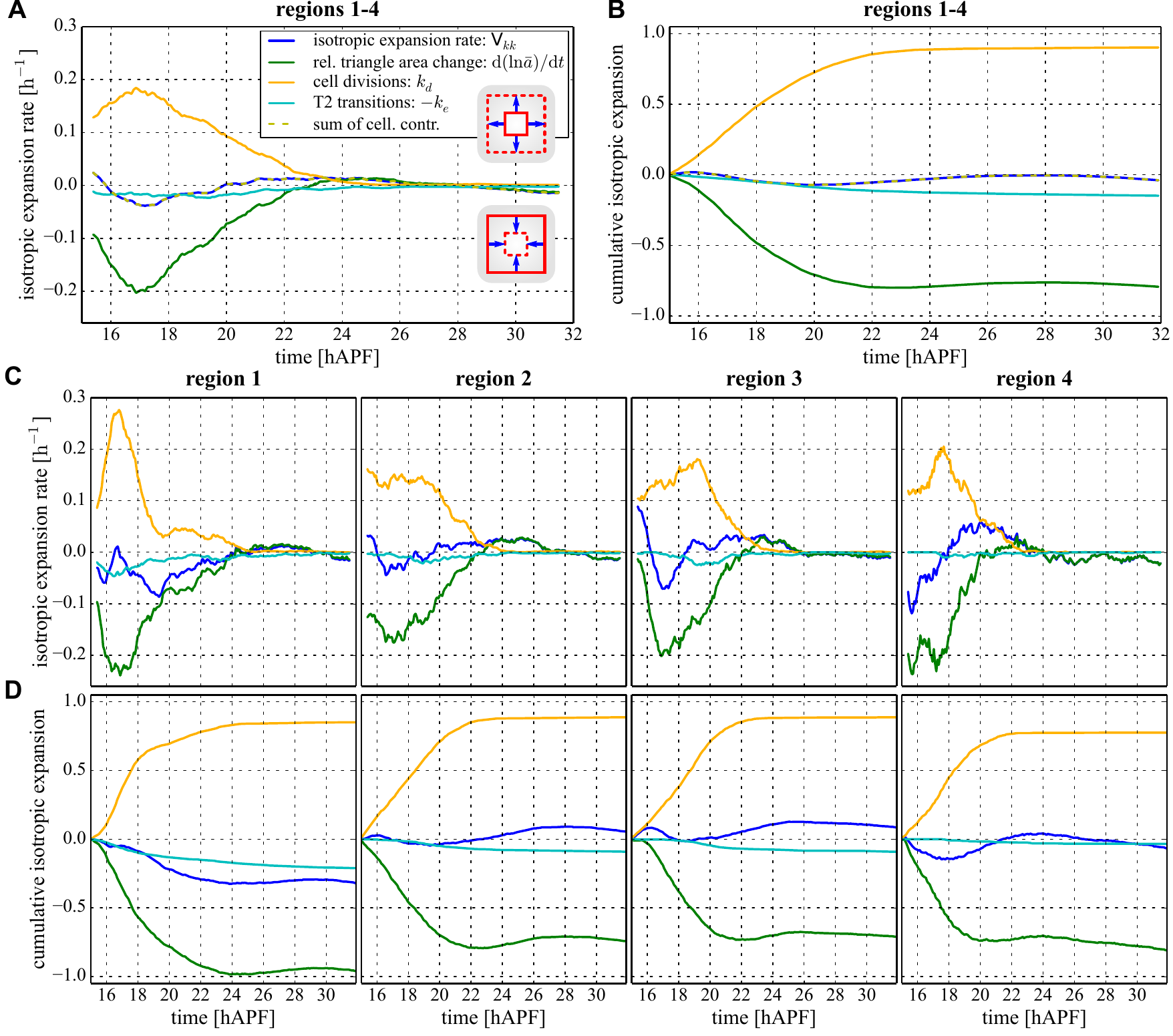}
  \caption{Contributions to isotropic tissue expansion as a function of time during pupal development of the fly wing. Shown is the data for one wing.
  (A) Large-scale isotropic expansion rate and cellular contributions to it averaged over regions 1-4 as a function of time.
  (B) Cumulative isotropic expansion rate and cellular contributions to it.
  (C,D) Same plots as in A and B, but for the subregions 1 to 4 indicated in \fref{fig:flyWing}A.
  The legend in panel A applies to panels B-D, too. In all panels, data was averaged over 10 subsequent inter-frame intervals.
  \label{fig:flyWing_iso}}
\end{figure*}
\subsection{Isotropic expansion rate}
We decompose the isotropic expansion rate $\itens{V}_{kk}$ as follows into cellular contributions:
\begin{equation}
  \itens{V}_{kk} = \frac{\d(\ln{\bar{a}})}{\d t} + k_d-k_e\text{.}\label{eq:decompositionIsotropicExpansion}
\end{equation}
Here, $\bar{a}$ is the average triangle area as in \sref{sec:largeScaleIsotropicExpansion}, and $k_d$ and $k_e$ denote cell division and cell extrusion rates, defined as
\begin{align}
 k_d &= \sum_{k\in\mathrm{CD}}{ \delta(t-t_k) \ln{\left(1+\frac{2}{N_k}\right)} }\label{eq:ieByCellDivisions} \\
 k_e &= -\sum_{k\in\mathrm{T2}}{ \delta(t-t_k) \ln{\left(1-\frac{2}{N_k}\right)} }\text{.}\label{eq:ieByT2Transitions}
\end{align}
The sums run over all topological transitions $k$ of the respective kind, $t_k$ denotes the time point of the respective transition, and $N_k$ is the number of triangles in the network before the respective transition.

Instead of formulating \eref{eq:decompositionIsotropicExpansion} for a triangulation, the polygonal network may also be used to derive such an equation. With the isotropic expansion rate for the polygonal network $\itens{V}_{kk}^p$, the average cell area $\bar{a}^p$ and the topological contributions by divisions $k_d^p$ and extrusions $k_e^p$, we obtain (see \aref{app:isotropicExpansionPolygonalNetwork}):
\begin{equation}
  \itens{V}_{kk}^p = \frac{\d(\ln{\bar{a}^p})}{\d t} + k_d^p-k_e^p\text{.}\label{eq:decompositionIsotropicExpansion_cells}
\end{equation}
This equation can be interpreted as a continuum equation for cell density \cite{Bittig2008,Ranft2010}, where the isotropic expansion rate contributions by cell divisions $k_d^p$ and cell extrusions $k_e^p$ correspond to cell division and cell extrusion rates, respectively.

\subsection{Cumulative shear and expansion}
Often, it is useful to consider cumulative deformations rather than deformation rates. The cumulative shear deformation is defined as $\int_{t_0}^{t_1}{\itens{\unem{V}}_{ij}\d t}$, other cumulative quantities are defined correspondingly. 
Note that this cumulative shear deformation is not a deformation that only depends on the initial and final configurations at times $t_0$ and $t_1$, but it also depends on the full path the system takes between those two configurations (see \aref{app:pathDependentPureShear}).
The cumulative isotropic expansion $\int_{t_0}^{t_1}{\itens{V}_{kk}\d t}=\log{A(t_1)} - \log{A(t_0)}$ is independent of the full path and given by a change of tissue area between initial and final states. This follows from \eref{eq:decompositionExpansionRateWithoutTopologicalTransitions}.
The cumulative shear can be decomposed into cellular contributions. This decomposition can be obtained by integrating the decomposition of shear rates \eref{eq:decompositionPureShear} over time. Similarly, the cumulative isotropic expansion can be decomposed into cellular contributions by integrating \eref{eq:decompositionIsotropicExpansion} over time.

\section{Tissue remodeling in the pupal fly wing as an example}
\label{sec:flyWing}
Our Triangle Method can be used to analyze tissue remodeling in the pupal fly wing \cite{Aigouy2010,Etournay2015}.
Here, we provide a more refined and in depth analysis of the wing morphogenesis data for three different wild type wings presented previously \cite{Etournay2015}.
Differences to the previous analyses are (i) there are slightly improved definitions of the shear rates for finite time intervals between frames (see \aref{app:quantificationSpatialAvgerages}) (ii) we now analyze and compare subregions of the wing tissue, which provides additional information about tissue remodeling.

\fref{fig:flyWing_patterns} presents coarse-grained spatial patterns of local tissue shear $\itens{\unem{V}}_{ij}$ (blue), the corotational time derivative of the cell elongation $\D\itens{\nem{Q}}_{ij}/\D t$ (green), and the contribution to shear by T1 transitions $\itens{\unem{V}}_{ij}$ (red) at different times during pupal development.
The bars indicate the local axis and strength of shear averaged in a small square.
The full dynamics of these patterns can be seen in the Movies~M1--M3.
Because here we do not track cells but use a lab frame relative to which the tissue moves, convective terms have been taken into account (see \aref{app:quantificationSpatialPatterns}).
The patterns in \fref{fig:flyWing_patterns} correspond to Figure~5 and Video~6 in ref.~\cite{Etournay2015}.
The pattern of tissue shear rate is splayed and decreases in magnitude over time.
The pronounced inhomogeneities of the shear pattern at $22\,\mathrm{hAPF}$ are due to different behaviors of veins and the intervein regions \cite{Etournay2016}. 
The orientations of the patterns of cell elongation change and shear by T1 transitions are both approximately homogeneous at early and late times.
At intermediate times, about $22\,\mathrm{hAPF}$, a reorientation of these patterns occurs, which corresponds to a transitions between a phase I and a phase II of tissue remodeling \cite{Aigouy2010,Etournay2015}.
During phase I, cells elongate along the proximal-distal axis of the wing while they are undergoing T1 transitions along the along the anterior-posterior axis of the wing.
During phase II, cells reduce their elongation along the proximal-distal axis while undergoing T1 transitions along this axis.

These dynamics and the two phases can be analyzed by averaging contributions to tissue shear in distinct subregions of the wing (see \fref{fig:flyWing}A) and in the whole wing blade.
We project the tensorial quantities on the $x$ axis, which is the average axis of cell elongation and is close to the proximal-distal axis (see \fref{fig:flyWing}B).
The quantities discussed are listed in \fref{fig:flyWing}C. 
The shear rates as a function of time and the corresponding cumulative shear are shown in \fref{fig:flyWing}D and E, respectively, averaged over the whole wing blade.
These data are consistent with the previous analysis \cite{Etournay2015}.
The fact that the sum of cellular contributions and tissue shear coincide in panels D and E confirms the validity of \eref{eq:decompositionPureShear} (blue and yellow dashed lines).

In panels F and G, we show shear rates and cumulative shear for the four subregions of the wing blade indicated in \fref{fig:flyWing}A and tracked in Movie~M4.
Comparing the average shear curves in \fref{fig:flyWing}F,G, we find systematic differences among the different regions. 
Most significantly, distal regions, which are regions closer to the tip of the wing (regions 3,4) shear more at early times, whereas proximal regions, i.e.\ regions closer to the hinge (regions 1,2), shear more towards the end of the process (blue curves).
Moreover, the cumulative shear at the end of the process is generally larger in distal regions than in proximal regions.
The transition from phase I to phase II can be seen in all four regions. However, it shifts from about $20.5\,\mathrm{hAPF}$ in region 4 to about $23\,\mathrm{hAPF}$ in region 1 (see for example intersection of red and green curves in panel F).
Finally, cell divisions contribute more to shear distally (region 4), whereas correlations effects contribute more to shear proximally (region 1).
All of these results, which we found consistently for the three analyzed wings, reveal a propagation of morphogenetic events through the tissue.

We also quantified the isotropic expansion rate $\itens{V}_{kk}$ and its cellular contributions, related by \eref{eq:decompositionIsotropicExpansion}. For the entire wing (\fref{fig:flyWing_iso}A,B), we again confirm our earlier results reported in \cite{Etournay2015}. We find that the total area of the wing blade barely changes (blue curve). Correspondingly, cell area decrease (green curve) together with contributions from cell extrusions (cyan curve) compensate most of the area changes due to cell divisions (orange curve).
When comparing the regions 1-4 (\fref{fig:flyWing_iso}C,D), area changes due to divisions occur earlier in region 1 and during a shorter time as compared to regions 2-4. Furthermore, region 1 does substantially shrink, whereas regions 2-4 barely change their areas. 
This difference may be related to the fact that the wing hinge contracts its area during this process.
All of these results are again consistent among the three analyzed wings.

\section{Discussion}
\label{sec:discussion}
In this article, we present a geometric analysis of tissue remodeling in two dimensions based on a triangulation of the cellular network.
We decompose the pure shear rate, the isotropic expansion rate, and the rotation rate of the tissue into cellular contributions.
The main result of this article is given by \eref{eq:decompositionPureShear}. It provides an exact expression of the large-scale shear rate as a sum of distinct cellular contributions, stemming from cell shape changes, T1 transitions, cell divisions, cell extrusions, and from correlation effects.
This decomposition is based on the fact that for a single triangle, shear deformations are related to cell elongation changes in a corotating reference frame, see \eref{eq:triangleShearRateAndQ}.
The corotating reference frame ensures that elongation changes associated with pure rotations do not give rise to shear deformations.
In the absence of rotations, small elongation changes and shear deformations are the same. 
Because of nonlinearities in the corotational time derivative, the average time derivative and the time derivative of the average differ (see \eref{eq:correlationTermIsCommutator}).
When coarse-graining, this gives rise to correlation contributions to tissue shear.
Such correlation terms exist when tissue remodeling is spatially inhomogeneous. 
For example, inhomogeneities of rotation rates give rise to correlation contributions to tissue shear that stem from correlations between rotation rates and triangle elongation (see \eref{eq:definitionCorrelationTerm}). Similarly, correlations between area changes and elongation also contribute to shear.
Thus, correlation contributions to large-scale tissue shear are a generic feature resulting from the interplay of nonlinearities and fluctuations. 

We have recently studied tissue morphogenesis in the pupal wing epithelium using our triangle method both in fixed reference frames and reference frames comoving with the tissue \cite{Etournay2015}. 
During pupal morphonesesis, the wing blade elongates along the proximal-distal axis while keeping its area approximately constant.
This process can be divided in two phases \cite{Aigouy2010}.
In the first phase, cells elongate more than the overall tissue does. This strong cell elongation is driven by active T1 transitions expanding perpendicular to the proximal-distal axis.
The cell elongation then subsequently relaxes during phase two by T1 transitions along the proximal-distal axis.
At late times, the tissue reaches a state with slightly elongated cells, which is a signature of active T1 transitions.
Also note that our analysis has shown that correlations contribute to tissue shear. In particular, we have shown that correlations between fluctuations of rotations and cell elongations occur and play a significant role for tissue morphogenesis. Our method can therefore detect biologically relevant processes that are otherwise difficult to spot. 

In the present article, we provide a refined analysis of these previously presented data, confirming our earlier findings. In addition, we perform a regional analysis of pupal wing remodeling.
Discussing the shear and cellular contributions to shear of the whole wing blade and in four different subregions, we find that the main morphogenetic processes of the wing \cite{Aigouy2010,Etournay2015} are also reflected in the different subregions.
However, the timing of these morphogenetic processes differs among the regions, revealing a propagation of morphogenetic events through the tissue.

Our work is related to other studies that decompose tissue shear into cell deformation and cell rearrangements \cite{Brodland2006,Graner2008,Blanchard2009,Kabla2010,Economou2013,Guirao2015}.
Our approach differs from these studies in that it provides an exact relation between cellular processes and tissue deformation gradients on all scales.
Recently, a method based on cell center connection lines rather than lines was presented \cite{Guirao2015}. This method is based on cell center connection lines rather than triangles.
While ref \cite{Guirao2015} and the methods presented here both provide a decomposition of shear into cellular contributions, the method presented here has an important property. 
We relate tissue deformations on all length scales to cellular contributions, taking into account correlation terms.
Simple area-weighted averaging of triangle-based quantities generates in our approach the corresponding coarse-grained quantities on large scales.

The Triangle Method described here provides a general framework to study tissue remodeling during morphogenesis.
We have focused our discussion on tissue deformations that are planar. It will be interesting to generalize our approach to curved surfaces and to bulk three-dimensional tissues. A generalization to three dimensions can be done following the same ideas and using tetrahedra. Almost all equations apply also in three dimensions, only \esref{eq:triangleShapeProperties} and \seref{eq:triangleShearAndQ} require special consideration of tetrahedral geometry.
Our approach can play an important role in understanding the complex rheology of cellular materials both living and non-living.

\section*{Acknowledgements}
This work was supported by the Max Planck Gesellschaft and by the BMBF.
MM also acknowledges funding from the Alfred P.\ Sloan Foundation, the Gordon and Betty Moore Foundation, and NSF-DMR-1352184.
RE acknowledges a Marie Curie fellowship from the 774 EU 7th Framework Programme (FP7). 
SE acknowledges funding from the ERC.

\appendix
\setcounter{secnumdepth}{2}

\section{Deformation of a triangle network}
\subsection{Deformation and deformation gradients}
\label{app:firstAppendix}
\label{app:genGauss}
For an Eucledian space, the following equation holds for a vector field $\vec{h}$:
\begin{equation}
  \int_\Lambda{\partial_i\ivec{h}_j\,\d A} = \oint_{\partial\Lambda}{\ivec{h}_j\ivec{\nu}_i\,\d\ell}\text{,}\label{eq:genGauss}
\end{equation}
where the area integral is over a domain $\Lambda$ with boundary $\partial\Lambda$. The vector $\vec{\nu}$ denotes the local unit vector that is normal to the boundary pointing outwards.

\eref{eq:genGauss} follows from Gauss' theorem:
\begin{equation}
  \int_\Lambda{\mathrm{div}\,\vec{a}\;\d A} = \oint_{\partial\Lambda}{\vec{a}\cdot\vec{\nu}\;\d\ell}\text{,}\label{eq:gauss}
\end{equation}
if the components of the vector $\vec{a}$ are chosen as
\begin{equation}
  \ivec{a}_k = \delta_{ik}\ivec{h}_j
\end{equation}
and $i,j$ are fixed.


\subsection{Triangulation of a cellular network}
\label{app:triangulation}
\subsubsection{Triangulation procedure}
Here, we define the triangulation procedure outlined in \sref{sec:triangulation} more precisely. An \textit{inner} vertex, i.e.\ a vertex that does not lie on the margin of the polygonal network, gives rise to one or several triangles. Any inner vertex touches at least three polygons.
An inner vertex that touches exactly three polygons $\alpha$, $\beta$, and $\gamma$ gives rise to a single triangle with corners $\vec{r}^\alpha$, $\vec{r}^\beta$, and $\vec{r}^\gamma$, as explained in \sref{sec:triangulation}.
Moreover, an inner vertex that touches $M$ with $M>3$ polygons $\alpha_1, \dots, \alpha_M$ gives rise to $M$ triangles, which are defined as follows. One corner of each of these $M$ triangles is defined by the average position $\vec{c}=(\alpha_1+\dots+\alpha_M)/M$. The other two corners of triangle $i$ with $1\leq i\leq M$ are defined by $\vec{r}^{\alpha_i}$ and $\vec{r}^{\alpha_{i+1}}$, where the index $i=M+1$ corresponds to the index $i=1$.

All non-inner vertices, i.e.\ those lying on the margin of the polygonal network, do not give rise to any triangles. As a result of that, a stripe along the margin of the polygonal network is not covered by triangles, which is ca.\ half a cell-diameter thick. 

Apart from this stripe, the resulting triangulation has no gaps between the triangles. Overlaps between the triangles are in principle possible. In such a case, at least one triangle can be assigned a negative area. However in our experimental data, such cases are very seldom.

\subsubsection{Effective number of triangles}
We compute the effective number $N$ of triangles a follows:
\begin{equation}
  N = \sum_{n\in V_{=3}}{1} + \sum_{n\in V_{>3}}{(M_n-2)}\text{.} 
\end{equation}
Here, $V_{=3}$ denotes the set of all inner three-fold vertices and $V_{>3}$ denotes the set of all inner $M$-fold vertices with $M>3$. The number $M_n$ is the number of cells touched by vertex $n$ (i.e.\ vertex $n$ is $M_n$-fold). Hence, all triangles arising from a three-fold vertex count as one effective triangle, and all $M$ triangles arising from a $M$-fold vertex with $M>3$ count as $(M-2)/M$ effective triangles.

An interpretation for this effective number $N$ of triangles is given by the following consideration.
An $M$-fold vertex with $M>3$ can be thought of as $M-2$ three-fold vertices that are so close to each other that they can not be distinguished from each other. If we transform each inner $M$-fold vertex with $M>3$ of our polygonal network into such $M-2$ three-fold vertices, then $N$ is the number of inner three fold-vertices in the resulting network. Put differently, $N$ is the number of triangles in the triangulation of the resulting network.

\subsection{Triangle shape}
\subsubsection{Side vectors of the reference triangle}
\label{app:referenceTriangleSides}
In a Cartesian coordinate system, the vectors $\vec{c}^\bondindex{\alpha\beta}$ describing the equilateral reference triangle are 
\begin{align}
\vec{c}^\bondindex{12}&=c_0\begin{pmatrix}1 \\ 0\end{pmatrix},
\\
\vec{c}^\bondindex{23}&=c_0\begin{pmatrix}-1/2 \\ \sqrt{3}/2\end{pmatrix},
\\
\vec{c}^\bondindex{31}&=c_0\begin{pmatrix}-1/2 \\ -\sqrt{3}/2\end{pmatrix}
\text{.}
\end{align}
Here, $c_0=2a_0^{1/2}/3^{1/4}$ is the side length and $a_0$ the area of the reference triangle.

\subsubsection{Extraction of shape properties from the triangle shape tensor}
\label{app:polarDecomposition}
Here, we show how to extract triangle area $a$, triangle elongation $\itens{\nem{q}}_{ij}$, and triangle orientation angle $\theta$ from the shape tensor $\itens{s}_{ij}$ according to \eref{eq:triangleShapeProperties}:
\begin{equation}
  \tens{s} = \left(\frac{a}{a_0}\right)^{1/2}\exp{(\tens{\nem{q}})}\cdot\tens{\rot}{\left(\theta\right)}\text{.}\label{eq:app:polarDecomposition}
\end{equation}
First, the area can be extracted by computing the determinant of this equation, which yields:
\begin{equation}
  a = a_0 \det{\tens{s}}\text{.}
\end{equation}
To compute $\itens{\nem{q}}_{ij}$ and $\theta$, it is useful to split the tensor $\itens{s}_{ij}$ into a symmetric, traceless part $\itens{\nem{s}}_{ij}$ and into a rest $\itens{h}_{ij}$ containing the trace and the antisymmetric part:
\begin{equation}
  \itens{s}_{ij} = \itens{\nem{s}}_{ij} + \itens{h}_{ij}\text{.}
\end{equation}
Then, the triangle orientation angle $\theta$ is such that $\itens{h}_{ij}$ corresponds to a rotation by $\theta$ up to a scalar factor $f$:
\begin{equation}
  \itens{h}_{ij} = f\itens{\rot}_{ij}{\left(\theta\right)}\text{,}
\end{equation}
and the triangle elongation can be computed as:
\begin{equation}
  \itens{\nem{q}}_{ij} = \frac{1}{\norm{\itens{\nem{s}}}}\ \mathrm{arcsinh}\,{\left[\left(\frac{a}{a_0}\right)^{-1/2}\norm{\itens{\nem{s}}}\right]}\itens{\nem{s}}_{ik}\itens{\rot}_{kj}{\left(-\theta\right)}\text{.}
\end{equation}
In \cite{Merkel2014b}, we show that these values for $a$, $\itens{\nem{q}}_{ij}$, and $\theta$ do indeed fulfill \eref{eq:app:polarDecomposition}, and that they are the unique solutions.

\subsubsection{Geometrical interpretation of the triangle elongation tensor}
\label{app:interpretationQ}
\begin{figure}
  \centering
  \includegraphics{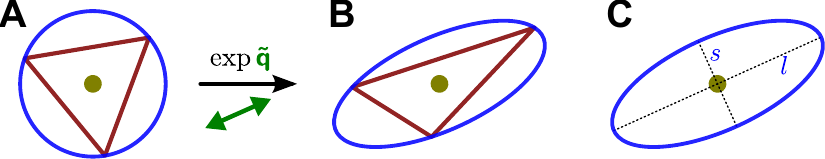}
  \caption{Geometrical interpretation of the elongation tensor $\itens{\nem{q}}_{ij}$ for a given triangle. (A) shows an equilateral triangle (red) with circumscribed circle (blue) and centroid (i.e.\ center of mass, yellow). (B) This triangle is deformed by the pure shear deformation given by $\exp{(\tens{\nem{q}})}$, where $\itens{\nem{q}}_{ij}$ is the elongation tensor of the so-created triangle. The former circumscribed circle is transformed to an ellipse (blue), and the former centroid is still the centroid of both the triangle and the ellipse (yellow). (C) Long and short axes of the ellipse with lengths $l$ and $s$, respectively.
  \label{fig:triangleShapeIllustration}}
\end{figure}
\fref{fig:triangleShapeIllustration} illustrates the geometrical interpretation of the triangle elongation tensor $\itens{\nem{q}}_{ij}$.
Take the unique ellipse (blue in \fref{fig:triangleShapeIllustration}B) that goes through all three corners of the triangle (red) and has the same center of mass (yellow) as the triangle. Then, the long axis of the ellipse corresponds to the axis of the triangle elongation tensor $\itens{\nem{q}}_{ij}$, and the aspect ratio of the ellipse is given by $l/s=\exp{(2\norm{\itens{\nem{q}}})}$ (\fref{fig:triangleShapeIllustration}C).

This can be seen as follows. As discussed in \sref{sec:triangleShape} and the previous section, any given triangle can be created out of an equilateral triangle using the pure shear transformation $\exp{(\tens{\nem{q}})}$, where $\itens{\nem{q}}_{ij}$ is the elongation tensor of the given triangle. This is illustrated in \fref{fig:triangleShapeIllustration}A-B. The circumscribed circle of the equilateral triangle transforms into the ellipse via the pure shear transformation. Thus, the length of the long and short axes of the ellipse are $l=r\exp{(\norm{\itens{\nem{q}}})}$ and $s=r\exp{(-\norm{\itens{\nem{q}}})}$, where $r$ is the radius of the circle.

The ellipse is uniquely defined, because the equilateral triangle and the pure shear deformation are uniquely defined as proven in \cite{Merkel2014b}. If there was another ellipse that went through all corners of the triangle and had the same center of mass, this ellipse could be created from a circle $c'$ using a different pure shear transformation. Applying the inverse of this pure shear transformation to the actual triangle $n$ would yield a triangle $n'$. Obviously, the triangle $n'$ would have the circumscribed circle $c'$ and thus its center of mass would coincide with the center of its circumscribed circle $c'$. Thus, $n'$ would be equilateral. However, this is not possible since there is only one equilateral triangle from which triangle $n$ can emerge by a pure shear deformation.


\subsection{Relation between triangle shape and triangle deformation}
\label{app:triangleshapeAndDeformationDerivation}
Here, we derive \esref{eq:triangleShearAndQ}-\seref{eq:triangleRotationAndTheta} in the main text. From \eref{eq:connectionTriangleShapeAndTransformation} follows with \eref{eq:triangleMu}:
\begin{equation}
  \itens{s}_{ij}^{\prime}-\itens{s}_{ij} = \itens{u}_{ki}\itens{s}_{kj}\label{eq:SandU}\text{.}
\end{equation}
For infinitesimal changes $\delta\itens{\nem{q}}_{ij}$, $\delta a$, $\delta\theta$ of the respective triangle shape properties, the difference of the shape tensors is also infinitesimal $\delta\itens{s}_{ij}=\itens{s}_{ij}^{\prime}-\itens{s}_{ij}$. From \eref{eq:app:polarDecomposition} follows:
\begin{equation}
  \begin{aligned}
    \delta\itens{s}_{ij} &= \frac{\delta a}{2a}\itens{s}_{ij} + \delta\norm{\itens{\nem{q}}}\frac{\itens{\nem{q}}_{ik}}{\norm{\itens{\nem{q}}}}\itens{s}_{kj} + \delta\phi\epsilon_{ik}\itens{s}_{kj} \\
    &\quad + (\delta\theta-\delta\phi)\itens{s}_{ik}\epsilon_{kj}\text{.}
  \end{aligned}
\end{equation}
Inserted into \eref{eq:SandU} and using the decomposition of the deformation tensor \eref{eq:decompositionU}, this yields:
\begin{equation}
  \begin{aligned}
  &\frac{1}{2}\delta\itens{u}_{kk}\delta_{ij} + \delta\itens{\unem{u}}_{ij} + \delta\psi\epsilon_{ij}  \\
  &= \frac{\delta a}{2a}\delta_{ij} + \delta\norm{\itens{\nem{q}}}\frac{\itens{\nem{q}}_{ij}}{\norm{\itens{\nem{q}}}} + \delta\phi\epsilon_{ij} \\
  &\quad+ (\delta\theta-\delta\phi)\itens{s}_{ik}\epsilon_{kl}\itens{s}^{-1}_{lj}\text{.}
  \end{aligned}
\end{equation}
To disentangle the contributions of the last term to the three deformation tensor components, we transform the tensor product into:
\begin{equation}
  \itens{s}_{ik}\epsilon_{kl}\itens{s}^{-1}_{lj} = \epsilon_{ik}\left[\cosh{(2\norm{\itens{\nem{q}}})}\delta_{kj} -\frac{\sinh{(2\norm{\itens{\nem{q}}})}}{\norm{\itens{\nem{q}}}}\itens{\nem{q}}_{kj}\right]\text{.}
\end{equation}
Hence, we obtain:
\begin{align}
  \delta\itens{\unem{u}}_{ij} &= \delta\norm{\itens{\nem{q}}}\frac{\itens{\nem{q}}_{ij}}{\norm{\itens{\nem{q}}}} - (\delta\theta-\delta\phi)\frac{\sinh{(2\norm{\itens{\nem{q}}})}}{\norm{\itens{\nem{q}}}}\epsilon_{ik}\itens{\nem{q}}_{kj}\label{eq:intermediateShearPart}\\
  \delta\itens{u}_{kk} &= \frac{\delta a}{a}\\
  \delta\psi &= \delta\phi + (\delta\theta-\delta\phi)\cosh{(2\norm{\itens{\nem{q}}})}\text{.}\label{eq:intermediateAnglePart}
\end{align}
\esref{eq:triangleShearAndQ}-\seref{eq:triangleRotationAndTheta} in the main text follow directly:
\begin{align}
  \delta\itens{\unem{u}}_{ij} &= \delta\itens{\nem{q}}_{ij} -2\Big[c\delta\psi+(1-c)\delta\phi\Big]\itens{\epsilon}_{ik}\itens{\nem{q}}_{kj}\label{eq:finalShearPart}\\
  \delta\itens{u}_{kk} &= \delta(\ln{a}) \\
  \delta\psi &= \delta\theta - \delta\itens{\unem{u}}_{ij}\itens{\epsilon}_{jk}\itens{\nem{q}}_{ki}\frac{\cosh{(2\norm{\itens{\nem{q}}})}-1}{2\norm{\itens{\nem{q}}}\sinh{(2\norm{\itens{\nem{q}}})}}\text{.}\label{eq:finalAnglePart}
\end{align}
with $c=\tanh{(2\norm{\itens{\nem{q}}})}/2\norm{\itens{\nem{q}}}$.
Here, to derive the expression for the pure shear part $\delta\itens{\unem{u}}_{ij}$, we used the decomposition of $\delta\itens{\nem{q}}_{ij}$ into contributions of norm and angle changes of $\itens{\nem{q}}_{ij}$, \eref{eq:changeOfQ}.
To derive the expression for the rotation part $\delta\psi$, we used that that from \eref{eq:intermediateShearPart} follows that:
\begin{equation}
  \delta\itens{\unem{u}}_{ij}\epsilon_{jk}\itens{\nem{q}}_{ki} = -2(\delta\theta-\delta\phi)\norm{\itens{\nem{q}}}\sinh{(2\norm{\itens{\nem{q}}})}\text{.}
\end{equation}

\subsubsection{Pure shear by triangle elongation change}
\label{app:corotationalDerivative}
To discuss the pure shear formula \eref{eq:finalShearPart}, we first consider the decomposition of an infinitesimal change of the triangle elongation tensor $\delta\itens{\nem{q}}_{ij}$ into a contribution by the change of the norm $\delta\norm{\itens{\nem{q}}}$ and a contribution by the change of the angle $\delta\phi$:
\begin{equation}
  \delta\itens{\nem{q}}_{ij} = \delta\norm{\itens{\nem{q}}}\frac{\itens{\nem{q}}_{ij}}{\norm{\itens{\nem{q}}}} +2\delta\phi\epsilon_{ik}\itens{\nem{q}}_{kj}\text{.}\label{eq:changeOfQ}
\end{equation}
The pure shear $\delta\itens{\unem{u}}_{ij}$ from \eref{eq:finalShearPart} can be rewritten in a similar form:
\begin{equation}
  \delta\itens{\unem{u}}_{ij} = \delta\norm{\itens{\nem{q}}}\frac{\itens{\nem{q}}_{ij}}{\norm{\itens{\nem{q}}}} +2c(\delta\phi-\delta\psi)\epsilon_{ik}\itens{\nem{q}}_{kj}\text{.}\label{eq:pureShearNormAndAngle}
\end{equation}
There are two differences between \eref{eq:changeOfQ} and \eref{eq:pureShearNormAndAngle} both of which affect the angular part. First, in \eref{eq:pureShearNormAndAngle}, the rotation $\delta\psi$ is subtracted from the angular change of the elongation tensor, $\delta\phi$. This accounts for bare rotations, which do change the elongation tensor $\itens{\nem{q}}_{ij}$ by changing its angle $\phi$, but do not contribute to pure shear $\delta\itens{\unem{u}}_{ij}$. Second, the ``rotation-corrected'' angle change of the elongation tensor, $\delta\phi-\delta\psi$, does not fully contribute to pure shear but is attenuated by a factor $c$ with $0< c\leq 1$, which depends nonlinearly on $\norm{\itens{\nem{q}}}$. This second point makes the corotational time derivative in \eref{eq:triangleMaterialDerivativeQ} different from other, more common time derivatives. However, for small elongations, $\norm{\itens{\nem{q}}}\ll 1$, we have $c\rightarrow 1$ and the corotational time derivative corresponds to the so-called Jaumann derivative \cite{Bird1987v2}.

\subsubsection{Shear-induced triangle rotation}
\label{app:shearInducedRotation}
Here, we discuss the shear-induced contribution $\delta\xi$ in \eref{eq:finalAnglePart}, which we rewrite as
\begin{equation}
  \delta\theta = \delta\psi + \delta\xi
\end{equation}
with
\begin{equation}
  \delta\xi = \delta\itens{\unem{u}}_{ij}\itens{\epsilon}_{jk}\itens{\nem{q}}_{ki}\frac{\cosh{(2\norm{\itens{\nem{q}}})}-1}{2\norm{\itens{\nem{q}}}\sinh{(2\norm{\itens{\nem{q}}})}}\label{eq:app_xi}\text{.}
\end{equation}
According to this equation, the triangle orientation angle $\theta$ may change even with vanishing $\delta\psi$ whenever there is pure shear that is neither parallel nor perpendicular to the elongation tensor $\itens{\nem{q}}_{ij}$, i.e.\ a pure shear that changes the elongation angle. 

\begin{figure}
  \centering
  \includegraphics{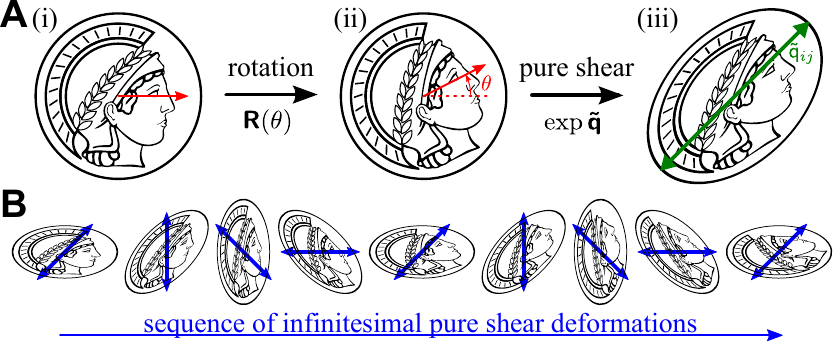}
  \caption{Illustration of the shear-induced rotation effect $\delta\xi$ appearing in \eref{eq:finalAnglePart}. For clarity, we use a Minerva head in place of a triangle. (A) The definitions of orientation angle $\theta$ and elongation tensor $\itens{\nem{q}}_{ij}$ are analogous to the triangle quantities (\fref{fig:triangleShapeNew}C). Roughly, the orientation angle $\theta$ corresponds to the direction in which the Minerva head looks. The isotropic scaling has been set to one for simplicity ($a=a_0$). (B) The elongated Minerva head is subject to a continuous pure shear deformation with varying shear axis. The pure shear axis is at each time point oriented with an angle of $\pi/4$ with respect to the elongation axis $\phi$, such that the elongation norm $\norm{\itens{\nem{q}}}$ does not change but only the angle $\phi$. Here, snapshots of such a deformation are shown. Alternatively, Movie~M5 shows this deformation more smoothly. Strikingly, the head orientation angle $\theta$ changes by this deformation although the deformation never includes any rotation component $\delta\psi=0$. Here, we have set $\norm{\itens{\nem{q}}}=(\ln{2})/2$ such that $\delta\theta=\delta\xi=0.2\delta\phi$.\label{fig:shearInducedRotation}}
\end{figure}
We illustrate this further in \fref{fig:shearInducedRotation}. For clarity, we use a Minerva head in place of a triangle, but with analogously defined shape and deformation properties (\fref{fig:shearInducedRotation}A). We discuss a continuous pure shear deformation of this head without rotation or isotropic expansion at any time point:
\begin{align}
  \delta\itens{u}_{kk} &= 0 \\
  \delta\psi &= 0 \text{.}
\end{align}
Because of the second equation, any potential change in the orientation angle $\theta$ must be due to the shear-induced effect: $\delta\theta=\delta\xi$.
Furthermore, the pure shear is defined such that the elongation norm $\norm{\itens{\nem{q}}}$ is constant, but the elongation angle $\phi$ may change. This can be accomplished by a pure shear axis that is at each time point at an angle of $\pi/4$ with respect to the elongation axis. This criterion can be written as:
\begin{equation}
  \delta\itens{\unem{u}}_{ij} =\delta h \epsilon_{ik}\itens{\nem{q}}_{kj}\text{,}
\end{equation}
where $\delta h$ is some infinitesimal scalar quantity. Comparison of this equation with \eref{eq:finalShearPart} and insertion into \eref{eq:app_xi} yields:
\begin{equation}
  \delta\xi = \delta\phi\left[1-\frac{1}{\cosh{(2\norm{\itens{\nem{q}}})}}\right]\text{.}
\end{equation}
Hence, although there is no rotation component of the deformation field $\delta\psi=0$, the orientation angle $\theta$ changes by a non-vanishing amount $\delta\theta=\delta\xi$ (\fref{fig:shearInducedRotation}B, Movie~M5).

\subsection{Large-scale pure shear}
\subsubsection{Relation to average elongation}
\label{app:largeScaleShear}
To find the relation between large-scale pure shear $\delta\itens{\unem{U}}_{ij}$ and large-scale elongation $\itens{\nem{Q}}_{ij}$, we average \eref{eq:triangleShearAndQ}:
\begin{equation}
  \delta\itens{\unem{U}}_{ij} = \big\langle\delta\itens{\nem{q}}_{ij}\big\rangle + \big\langle\delta\itens{\nem{j}}_{ij}\big\rangle\text{.}
\end{equation}
To show \eref{eq:decompositionShearWithoutTopologicalTransitions}, it remains to be shown that:
\begin{equation}
  \delta\itens{\nem{Q}}_{ij} =  \big\langle\delta\itens{\nem{q}}_{ij}\big\rangle + \big\langle\delta\itens{u}_{kk}\itens{\nem{q}}_{ij}\big\rangle-  \delta\itens{U}_{kk}\itens{\nem{Q}}_{ij}\text{.}
\end{equation}
This equation reflects the fact that changes in the triangle areas also contribute to a change in the average elongation $\itens{\nem{Q}}_{ij}$. Formally, the equation can be derived using the definition of the average elongation $\itens{\nem{Q}}_{ij}=\langle\itens{\nem{q}}_{ij}\rangle$ together with \esref{eq:triangleGAndA} and \seref{eq:decompositionExpansionWithoutTopologicalTransitions}.

\subsubsection{Correlation terms arising from convective and corotational terms}
\label{app:correlationConvection}
Here, we show how the correlation term $\itens{\nem{D}}_{ij}$ arises from convective and corotational terms. To this end, we introduce continuous, time-dependent fields for shear rate $\unem{v}_{ij}(\vec{r}, t)$ and triangle elongation $\nem{Q}_{ij}(\vec{r}, t)$. Whenever a given position $\vec{r}$ lies inside of a triangle $n$ at time point $t$, both are defined by:
\begin{align}
  \unem{v}_{ij}(\vec{r}, t) &= \itens{\unem{v}}_{ij}^n \\
  \nem{Q}_{ij}(\vec{r}, t) &= \itens{\nem{q}}_{ij}^n\text{.}
\end{align}
Given these definitions, \eref{eq:triangleShearRateAndQ} can be rewritten as:
\begin{equation}
  \unem{v}_{ij} = \frac{\mathcal{D}\nem{Q}_{ij}(\vec{r}, t)}{\mathcal{D}t}\label{eq:app:shearRateQ}
\end{equation}
with the corotational time derivative $\mathcal{D}\nem{Q}_{ij}/\mathcal{D}t$ defined as
\begin{equation}
  \frac{\mathcal{D}\nem{Q}_{ij}}{\mathcal{D}t} = \frac{\partial\nem{Q}_{ij}(\vec{r}, t)}{\partial t}  + \ivec{\mathrm{v}}_k\partial_k\nem{Q}_{ij} + \frac{\delta\itens{\nem{j}}_{ij}^n}{\delta t}\text{.}\label{eq:app:cotorDerivative}
\end{equation}
Here, $n$ is the triangle which contains the position $\vec{r}$ at time $t$. The vector $\ivec{\mathrm{v}}_k$ denotes the velocity field that is obtained by linear interpolation between the cell center velocities, i.e.\ by $\vec{\mathrm{v}}(\vec{r})\delta t=\vec{h}(\vec{r})$ with $\vec{h}(\vec{r})$ given by \eref{eq:triangleDefinitionU}.

In \eref{eq:app:cotorDerivative}, we take the corotational term $\delta\itens{\nem{j}}_{ij}^n/\delta t$ directly from the triangle-related \eref{eq:triangleShearRateAndQ}. However in addition, a convective term $\ivec{\mathrm{v}}_k\partial_k\nem{Q}_{ij}$ needs to be introduced for the following reason. The partial time derivative $\partial\nem{Q}_{ij}(\vec{r}, t)/\partial t$ on the right-hand side is essentially different from the ``total'' time derivative $\delta\itens{\nem{q}}_{ij}^n/\delta t$ appearing in \eref{eq:triangleMaterialDerivativeQ}: Whenever the tissue moves such that the boundary between two triangles passes past position $\vec{r}$, the partial time derivative contains a Dirac $\delta$ peak, which is not contained in the ``total'' time derivative. This peak is exactly compensated for by the convective term, which is only nonzero at triangle boundaries.

To obtain the large-scale shear rate of a triangulation, we can coarse-grain \eref{eq:app:shearRateQ} instead of the triangle relation \eref{eq:triangleShearRateAndQ}. Eventually, we should obtain the same relation for the large-scale shear rate, \eref{eq:decompositionShearRateWithoutTopologicalTransitions}. By comparing both ways, we can spot which term in the continuum formulation give rise to which terms in the triangle formulation.

To coarse grain \eref{eq:app:shearRateQ}, we write the large-scale shear rate $\itens{\unem{V}}_{ij}$ as follows (using \eref{eq:UOmegaContinuousArea}):
\begin{equation}
  \itens{\unem{V}}_{ij} = \langle\unem{v}_{ij}\rangle\text{,}\label{eq:app:largeScaleShearRate}
\end{equation}
where the averaging bracket is defined as follows
\begin{equation}
  \langle\unem{v}_{ij}\rangle = \frac{1}{A}\int_{\Lambda}{\unem{v}_{ij}\,\d A}\text{.}
\end{equation}
Here, the integration is over the whole triangle network $\Lambda$ with area $A$.
Substituting \eref{eq:app:shearRateQ} into \eref{eq:app:largeScaleShearRate} yields:
\begin{equation}
  \begin{aligned}
    \itens{\unem{V}}_{ij} &= \left\langle\frac{\partial\nem{Q}_{ij}(\vec{r}, t)}{\partial t}\right\rangle
    - \Big\langle(\partial_k\ivec{\mathrm{v}}_k)\nem{Q}_{ij}\Big\rangle + \left\langle\frac{\delta\itens{\nem{j}}_{ij}}{\delta t}\right\rangle\\
    &\qquad + \frac{1}{A}\oint_{\partial\Lambda}{\ivec{\nu}_k\ivec{\mathrm{v}}_k\nem{Q}_{ij}\,\d A}\text{.}\label{eq:app:correlationsAppearing}
  \end{aligned}
\end{equation}
Here, we carried out a partial integration on the term arising from the convective term, which gave rise to the boundary integral. In the boundary integral, the vector $\ivec{\nu}_k$ denotes the unit vector normal to the boundary, pointing outwards.

The second and the third terms in \eref{eq:app:correlationsAppearing} are essential parts of the correlation term $\itens{\nem{D}}_{ij}$. In particular, the term $-\langle(\partial_k\ivec{\mathrm{v}}_k)\nem{Q}_{ij}\rangle=-\langle\itens{v}_{kk}\itens{\nem{q}}_{ij}\rangle$, which arose from the convective term, is an essential part of the growth correlation. Similarly, the term $\langle\delta\itens{\nem{j}}_{ij}/\delta t\rangle$ is an essential part of the rotational correlation. 

To obtain \eref{eq:decompositionShearRateWithoutTopologicalTransitions} from \eref{eq:app:correlationsAppearing}, we note that the average elongation is $\itens{\nem{Q}}_{ij} = \langle\nem{Q}_{ij}\rangle$, and transform its total time derivative:
\begin{equation}
  \begin{aligned}
    \frac{\delta\itens{\nem{Q}}_{ij}}{\delta t} &= \frac{1}{\delta t}\left(\frac{1}{A(t+\delta t)}-\frac{1}{A(t)}\right)\int_{\Lambda(t)}{\nem{Q}_{ij}\,\d A} \\
    &\quad+\frac{1}{A\delta t}\left(\int_{\Lambda(t+\delta t)}{\nem{Q}_{ij}\,\d A}-\int_{\Lambda(t)}{\nem{Q}_{ij}\,\d A}\right)\\
    &\quad+\frac{1}{A\delta t}\int_{\Lambda(t)}{\delta\nem{Q}_{ij}\,\d A}\text{.}
  \end{aligned}
\end{equation}
These three terms can be respectively transformed into:
\begin{equation}
  \frac{\delta\itens{\nem{Q}}_{ij}}{\delta t} = -\itens{V}_{kk}\itens{\nem{Q}}_{ij} + \frac{1}{A}\int_{\partial\Lambda}{\ivec{\nu}_k\ivec{\mathrm{v}}_k\nem{Q}_{ij}\,\d A} + \left\langle\frac{\partial\nem{Q}_{ij}(\vec{r}, t)}{\partial t}\right\rangle\text{.}\label{eq:app:totalDerivativeQ}
\end{equation}
The first term is the mean-field term in the growth correlation and the second term is the boundary term generated by the convective term. Both terms appear due to a possible change of the triangulation domain $\Lambda$.
After all, \eref{eq:decompositionShearRateWithoutTopologicalTransitions} follows by inserting \eref{eq:app:totalDerivativeQ} into \eref{eq:app:correlationsAppearing}.

\subsection{Pure shear by a single T1 transition}
\label{app:T1shear}
A more general idea to define the pure shear by a single T1 transition could be the following. One could virtually deform triangles $m$ and $n$ in \fref{fig:topologicalTransitions}A to coincide with the shapes of the triangles $p$ and $q$, respectively. During such a virtual deformation, we would allow gaps and overlaps between triangles. Then one could set the shear by the T1 transition $\Delta\itens{\nem{X}}_{ij}$ to the shear rate integrated throughout this virtual deformation. However, as we point out in \aref{app:pathDependentPureShear}, such an integrated shear does not only depend on initial and final state, but also on the precise way the deformation is carried out. In that sense, there is no ``natural'' definition for $\Delta\itens{\nem{X}}_{ij}$ and a choice needs to be made.

The proposed definition $\Delta\itens{\nem{X}}_{ij}=-\Delta\itens{\nem{Q}}_{ij}$ corresponds to the following virtual deformation path of the triangles $m$ and $n$ into the shapes of the triangles $p$ and $q$. The initial triangles $m$ and $n$ are first separately deformed by pure shear deformations perpendicular to their respective elongation axes such that their elongations become zero. Then, both are rotated and scaled to assume the absolute orientation angles $\theta$ and areas of triangles $p$ and $q$, respectively. Finally, both triangles again undergo separate pure shear deformations to obtain the elongations of $p$ and $q$, respectively. Evaluating the integrated average pure shear rate $\itens{\unem{V}}_{ij}$ throughout this deformation yields indeed $-\Delta\itens{\nem{Q}}_{ij}$.

Moreover, the corotational term $\delta\itens{\nem{J}}_{ij}$ and the correlations $\delta\itens{\nem{K}}_{ij}$ never contribute to pure shear during this virtual deformation. Correspondingly, we set the ``overall'' $\delta\itens{\nem{J}}_{ij}$ and $\delta\itens{\nem{K}}_{ij}$ to zero at the time point of the T1 transition. Note however, that although $\delta\itens{\nem{J}}_{ij}=0$, the time derivative of the elongation angle $\Phi$ enters in the definition of $\delta\itens{\nem{J}}_{ij}$. But $\Phi$ typically changes by a finite amount during the T1 transition. Our choice of defining $\delta\itens{\nem{J}}_{ij}=0$ during a T1 transition thus implies to ignore the resulting Dirac $\delta$ peak.

\subsubsection{Special case: square or rhombus}
\label{app:T1shear:rhombus}
Here, we derive the shear by a single T1 transition for the special case where the four involved cell centers (green dots in \fref{fig:T1transitionQ}) form a square or, more generally, a rhombus.

For the case of a square, all involved triangles are isosceles triangles with a base angle of $\pi/4$. Such a triangle has an elongation tensor with an axis parallel to the base and with the norm $\norm{\itens{\nem{q}}^n}=(\ln{3})/4$. This can be shown using the formulas presented in \aref{app:polarDecomposition}, or by the following reasoning.
We ask for the shape tensor $\itens{s}^n_{ij}$ needed to transform an equilateral reference triangle into an isosceles triangle with the same area and a base angle of $\pi/4$. We set one of the sides of the reference triangle and the base of the isosceles triangle parallel to the $x$ axis. Then, the ratio of the base length of the isosceles triangle to the side length of the reference triangle is $3^{1/4}$, and the ratio of the heights of both triangles is $3^{-1/4}$. Correspondingly, the shape tensor reads
\begin{equation}
  \tens{s}^n = \begin{pmatrix} 3^{1/4} & 0 \\ 0 & 3^{-1/4}\end{pmatrix}\text{.}
\end{equation}
This shape tensor corresponds to the elongation tensor $\itens{\nem{q}}^n_{ij}$ that is parallel to the $x$ axis and has norm $\norm{\itens{\nem{q}}^n}=(\ln{3})/4$.

The shear by the T1 transition is given by the change of the average elongation tensor. For the case of a square, both triangles before and after the T1 transition have the same elongation tensor with norm $\norm{\itens{\nem{q}}^n}=(\ln{3})/4$. Thus, also the average elongation tensors for the square before and after the T1 transition have norm $\norm{\itens{\nem{Q}}}=(\ln{3})/4$. However, the axes of both average elongation tensors are perpendicular to each other, oriented along the diagonals of the square. Thus the shear by the T1 transition, which is given by the change of the average elongation tensor has norm $\norm{\Delta\nem{X}}=(\ln{3})/2$. 

The more general case of a rhombus can be treated by transforming the rhombus into a square by a pure shear transformation along the short diagonal of the rhombus. The effects of this pure shear transformation on the average elongation tensors before and after the T1 transition cancel out exactly. Note however that this argument only works, because the axis of this pure shear transformation is parallel or perpendicular to the elongation axes of all involved triangles.

In the above arguments, the average elongation was computed only for the rhombus with area $A_\square$. However, when the triangulation under consideration extends beyond the rhombus and has area $A$, the norm of the shear by the T1 transition results to be $\norm{\Delta\nem{X}}=(A_\square\ln{3})/(2A)$.

\subsection{Cellular contributions to isotropic expansion of a polygonal network}
\label{app:isotropicExpansionPolygonalNetwork}
We derive a decomposition of the isotropic expansion rate $\itens{V}_{kk}^p$ of the polygonal network. To this end, we first define the infinitesimal deformation tensor $\delta\itens{U}_{ij}^p$ for the whole polygonal network using a variant of \eref{eq:UOmegaTrianglesBoundary}, where we sum over polygon edges $b$ along the outline of the polygonal network instead of triangle sides along the outline of the triangular network:
\begin{equation}
  \delta\itens{U}_{ij}^p = \frac{1}{A^p}\sum_{b}{\delta\ivec{h}_j^{b}\ivec{\nu}_i^{b}\Delta\ell^{b}}\text{.}
\end{equation}
Here, $A^p$ is the area of the polygonal network, the vector $\ivec{\nu}_i^{b}$ is the unit vector normal to side $b$ that points outside, the scalar $\Delta\ell^{b}$ is the length of side $b$, and $\delta\ivec{h}_j^{b}=(\delta\ivec{h}_j^{m}+\delta\ivec{h}_j^{n})/2$ with $m$ and $n$ being the vertices at the ends of edge $b$, and $\delta\ivec{h}_j^{m}$ and $\delta\ivec{h}_j^{n}$ being their respective displacement vectors.

Then we have that:
\begin{equation}
  \delta\itens{U}_{kk}^p = \frac{\delta A^p}{A^p}\text{,}
\end{equation}
where $\delta A^p$ is the change of the area across the deformation. This equation can be shown using that $A^p=\sum_{b}{\ivec{r}_k^{b}\ivec{\nu}_k^{b}\Delta\ell^{b}}$ where the sum is over all polygon edges $b$ along the outline of the polygonal network, $\ivec{r}_k^{b}=(\ivec{r}_k^{m}+\ivec{r}_k^{n})/2$ with $m$ and $n$ being the vertices at the ends of edge $b$, and $\ivec{r}_k^{m}$ and $\ivec{r}_k^{n}$ being their respective positions.

Defining the average cell area by $\bar{a}^p=A^p/N^p$ where $N^p$ is the number of cells in the polygonal network, we have for the case without topological transitions:
\begin{equation}
  \delta\itens{U}_{kk}^p = \delta(\ln{\bar{a}^p})\text{.}
\end{equation}
Topological transitions are accounted for as explained in \sref{sec:isotropicExpansionSingleTt}. Hence, we finally obtain \eref{eq:decompositionIsotropicExpansion_cells} with 
\begin{align}
 k_d^p &= \sum_{k\in\mathrm{CD}}{ \delta(t-t_k) \ln{\left(1+\frac{1}{N_k^p}\right)} } \\
 k_e^p &= -\sum_{k\in\mathrm{T2}}{ \delta(t-t_k) \ln{\left(1-\frac{1}{N_k^p}\right)} }\text{,}
\end{align}
where the sums run over all topological transitions $k$ of the respective kind, $t_k$ denotes the time point of the respective transition, and $N_k^p$ is the number of cells in the network before the respective transition.

\subsection{Cellular contributions to large-scale rotation in a triangle network}
\label{app:rotationTriangleNetwork}
For the sake of completeness, we discuss the decomposition of large-scale rotation $\Omega=\langle\omega\rangle$, i.e.\ $\Omega\delta t=\delta\Psi$, into cellular contributions similar to the shear rate decomposition \eref{eq:decompositionPureShear}. In particular, we want to relate $\Omega$ to average triangle orientation, which we characterize using the complex hexatic order parameter $P_6$ with:
\begin{equation}
  P_6 = \langle p_6\rangle \qquad\text{with}\qquad p_6=e^{6i\theta}\text{.}
\end{equation}
Here, we use again an area-weighted average over all triangles, $i$ denotes the imaginary unit, and $\theta$ is the triangle orientation angle defined in \eref{eq:triangleShapeProperties}.

In the absence of topological transitions, the change of the hexatic order parameter $P_6$ relates to the large-scale rotation rate $\Omega$ as follows (using \eref{eq:triangleVorticity}):
\begin{equation}
  \begin{aligned}
    \frac{\d P_6}{\d t} &= 
      6iP_6\bigg[\Omega  + \itens{\unem{V}}_{ij}\itens{\epsilon}_{jk}\itens{\nem{Q}}_{ki}f(\norm{\itens{\nem{Q}}})\bigg] \\
      &\quad + \Big(\langle \itens{v}_{kk}p_6\rangle - \itens{V}_{kk}P_6\Big) + 6i\Big(\langle \omega p_6\rangle - \Omega P_6\Big) \\
      &\quad + 6i\bigg(\Big\langle \itens{\unem{v}}_{ij}\itens{\epsilon}_{jk}\itens{\nem{q}}_{ki}f(\norm{\itens{\nem{q}}})p_6\Big\rangle  - \itens{\unem{V}}_{ij}\itens{\epsilon}_{jk}\itens{\nem{Q}}_{ki}f(\norm{\itens{\nem{Q}}})P_6\bigg) \label{eq:P6Omega}
  \end{aligned}
\end{equation}
with $f(w)=[\cosh{(2w)}-1]/[2w\sinh{(2w)}]$.

The complex hexatic order parameter $P_6$ contains two pieces of information, the magnitude $Z_6$ of hexatic order and its orientation $\Theta_6$, which are real numbers defined by:
\begin{equation}
  P_6 = Z_6e^{6i\Theta_6}\text{.}\label{eq:P6Decomposition}
\end{equation}
Here, the orientation angle is defined to lie within the interval $-\pi/6<\Theta_6\leq\pi/6$. The value of the magnitude can be expressed as the average $Z_6=\langle\cos{(6[\theta-\Theta_6])}\rangle$.
Using \eref{eq:P6Decomposition}, \eref{eq:P6Omega} splits into an equation for the magnitude:
\begin{equation}
  \begin{aligned}
    \frac{\d Z_6}{\d t} &= \Big\langle\itens{v}_{kk}\cos{(6[\theta-\Theta_6])}\Big\rangle - \itens{V}_{kk}Z_6 \\
    &\quad -6\Big\langle\omega\sin{(6[\theta-\Theta_6])}\Big\rangle \\
    &\quad -6\Big\langle \itens{\unem{v}}_{ij}\itens{\epsilon}_{jk}\itens{\nem{q}}_{ki}f(\norm{\itens{\nem{q}}})\sin{(6[\theta-\Theta_6])}\Big\rangle
  \end{aligned}
\end{equation}
and into an equation characterizing the orientation:
\begin{equation}
  \frac{\d \Theta_6}{\d t} = \Omega + \itens{\unem{V}}_{ij}\itens{\epsilon}_{jk}\itens{\nem{Q}}_{ki}f(\norm{\itens{\nem{Q}}}) + \Sigma \label{eq:Theta6Omega}
\end{equation}
with correlations $\Sigma$ given by:
\begin{equation}
  \begin{aligned}
    \Sigma &= \frac{1}{Z_6}\Bigg[\frac{1}{6}\Big\langle\itens{v}_{kk}\sin{(6[\theta-\Theta_6])}\Big\rangle \\
      &\qquad\quad + \bigg(\Big\langle\omega\cos{(6[\theta-\Theta_6])}\Big\rangle - \Omega Z_6\bigg) \\
      &\qquad\quad + \bigg(\Big\langle\itens{\unem{v}}_{ij}\itens{\epsilon}_{jk}\itens{\nem{q}}_{ki}f(\norm{\itens{\nem{q}}})\cos{(6[\theta-\Theta_6])}\Big\rangle \\
      &\qquad\qquad\qquad\qquad\quad - \itens{\unem{V}}_{ij}\itens{\epsilon}_{jk}\itens{\nem{Q}}_{ki}f(\norm{\itens{\nem{Q}}}) Z_6\bigg)\Bigg]
    \text{.}
  \end{aligned}
\end{equation}
\eref{eq:Theta6Omega} relates the orientation of the hexatic order $\Theta_6$, which can be interpreted as an average triangle orientation, to the large-scale vorticity $\Omega$. For what follows, we multiply \eref{eq:Theta6Omega} with $\delta t$:
\begin{equation}
  \delta\Psi = \delta\Theta_6 - \delta\itens{\unem{U}}_{ij}\itens{\epsilon}_{jk}\itens{\nem{Q}}_{ki}f(\norm{\itens{\nem{Q}}}) - \Sigma\delta t\text{.}\label{eq:deltaPsiDeltaTheta6}
\end{equation}
Here, $\delta\Theta_6$ denotes the change of the average triangle orientation $\Theta_6$. Note the analogy of this equation with \esref{eq:decompositionShearWithoutTopologicalTransitions} and \seref{eq:decompositionExpansionWithoutTopologicalTransitions}.

To account for the effect of topological transitions, one can proceed as in \sref{sec:ttcontributionsToDeformation}. 
The displacement gradient across a topological transition is zero and so is its anisotropic part $\Delta\Psi=0$ and the shear $\Delta\itens{\nem{U}}_{ij}=0$. To account for example for a T1 transition, we introduce a new term $\Delta\Xi_6^T$ into \eref{eq:deltaPsiDeltaTheta6}, which represents the rotation by the T1 transition:
\begin{equation}
  \Delta\Psi = \Delta\Theta_6 - \Delta\itens{\unem{U}}_{ij}\itens{\epsilon}_{jk}\itens{\nem{Q}}_{ki}f(\norm{\itens{\nem{Q}}}) + \Delta\Xi_6^T\text{.}\label{eq:deltaPsiDeltaTheta6T1}
\end{equation}
Here, $\Delta\Theta_6$ is the change of $\Theta_6$ induced by the T1 transition, and we have set the correlations across the T1 transition to zero as we did in \sref{sec:ttcontributionsToShear}. After all, we obtain from \eref{eq:deltaPsiDeltaTheta6T1} that $\Delta\Xi_6^T=-\Delta\Theta_6$. 

Wrapping up, we find the following decomposition of the large-scale vorticity:
\begin{equation}
  \Omega = \frac{\d \Theta_6}{\d t} - \itens{\unem{V}}_{ij}\itens{\epsilon}_{jk}\itens{\nem{Q}}_{ki}f(\norm{\itens{\nem{Q}}}) + \Gamma_6^T + \Gamma_6^C + \Gamma_6^E - \Sigma
\end{equation}
with the rotations by T1 transitions $\Gamma_6^T$, cell divisions $\Gamma_6^C$, and cell extrusions $\Gamma_6^E$ defined by:
\begin{align}
  \Gamma_6^T &= - \sum_{k\in\mathrm{T1}}{\Delta\Theta_6^k\,\delta(t-t_k)} \\
  \Gamma_6^C &= - \sum_{k\in\mathrm{CD}}{\Delta\Theta_6^k\,\delta(t-t_k)} \\
  \Gamma_6^E &= - \sum_{k\in\mathrm{T2}}{\Delta\Theta_6^k\,\delta(t-t_k)}\text{.}
\end{align}
Here, the sums run over all topological transitions $k$ of the respective kind, $t_k$ denotes the time point of the respective transition, and $\Delta\Theta_6^k$ denotes the instantaneous change in $\Theta_6$ induced by the transition.

Note that in principle, one could also use for instance the triatic order parameter:
\begin{equation}
  P_3 = \Big\langle e^{3i\theta}\Big\rangle\text{.}
\end{equation}
However, for our purposes we prefer to use $P_6$ over $P_3$. This is because for a regular hexagonal array of cells, $P_3$ vanishes, whereas $P_6$ is nonzero. Hence, $P_6$ would allow us to track large-scale rotations of a regular hexagonal pattern of cells, which would not be possible using $P_3$.

\subsection{Path-dependence of the cumulative pure shear}
\label{app:pathDependentPureShear}
Here, we discuss the finite deformation of a triangular network that starts from a state with configuration $I$ and ends in another state with configuration $F$.  The initial and final configuration $I$ and $F$ respectively define all triangle corner positions and the topology of the network.
We define the corresponding cumulative pure shear by:
\begin{equation}
  \int_I^F{\delta\itens{\unem{U}}_{ij}} = \int_0^T{\itens{\unem{V}}_{ij}\,\d t}\text{,}
\end{equation}
where the deformation starts at time $0$ in state $I$ and ends at time $T$ in state $F$.

\begin{figure}
  \centering
  \includegraphics{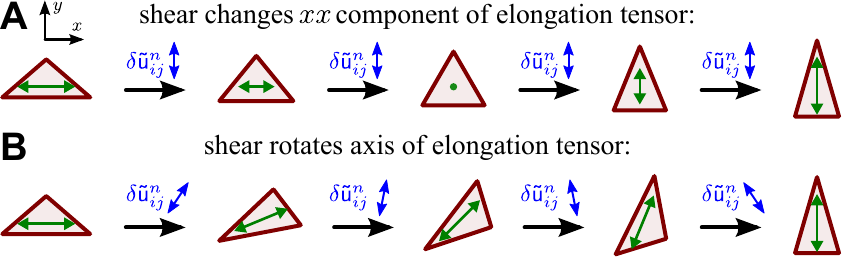}
  \caption{Path-dependence of the cumulative pure shear. Shown are two finite deformations with the same initial and final states, but with a different cumulative shear. Initial and final states are isosceles triangles, with the same elongation norm. (A) The triangle is sheared along the $y$ axis. (B) The triangle is sheared such that the elongation tensor is rotated but the norm stays constant.\label{fig:pathDependentIntegratedShear}}
\end{figure}
The cumulative pure shear does not only depend on the initial and final states $I$ and $F$, but also on the network states in between. We demonstrate this path-dependence of the cumulative pure shear for the case of a single triangle (\fref{fig:pathDependentIntegratedShear}). The initial state $I$ is given by a triangle with an elongation tensor parallel to the $x$ axis with $\itens{\nem{q}}^n_{xx}=Q_0$, where $Q_0$ is a positive scalar. The final state $F$ is given by a triangle with an elongation tensor parallel to the $y$ axis with $\itens{\nem{q}}^n_{xx}=-Q_0$. In initial and final states, the triangle areas are the same and in both states, $\theta^n=0$. \fref{fig:pathDependentIntegratedShear} illustrates two different deformation paths to reach state $F$ from state $I$. In \fref{fig:pathDependentIntegratedShear}A, the triangle is sheared along the horizontal, which corresponds to an cumulative shear of:
\begin{equation}
  \int_{\text{A}}{\delta\itens{\unem{u}}_{ij}^n} = -2Q_0\text{.}
\end{equation}
This follows from \eref{eq:triangleShearAndQ}.
In \fref{fig:pathDependentIntegratedShear}B, the triangle is undergoes a time-dependent pure shear such that the elongation axis is rotated but its norm stays constant. At the same time, to ensure that the orientation angle does not change $\delta\theta^n=0$, the rotation $\delta\psi^n$ as given by \eref{eq:triangleRotationAndTheta} is nonzero. The additional contributions by the corotational term in \eref{eq:triangleShearAndQ} eventually yield \citep{Merkel2014b}:
\begin{equation}
  \int_{\text{B}}{\delta\itens{\unem{u}}_{ij}^n} = -\sinh{(2Q_0)}\text{.}
\end{equation}
Thus, the cumulative pure shear for both integration paths is different -- or put differently, the cumulative shear is path-dependent.
Note that an equivalent statement to the path-dependence of pure shear is that the cumulative shear over a cyclic deformation is in general nonzero, where by \textit{cyclic} deformation, we mean a deformation with coinciding initial and the final states. 

Finally, we remark that at least for a triangular network with more than two triangles, the path-dependence of the cumulative pure shear can be generalized as follows \citep{Merkel2014b}. We consider a set of tensors $G_{ij}^\mathrm{tr}$, $G_{ijkl}^{s}$, $G_{ij}^\mathrm{a}$, and $H_{ij}$ that only depend on the given state of the triangular network. Then, the following equation:
\begin{equation}
  \int_I^F{\Big(G_{ij}^\mathrm{tr}\delta\itens{U}_{kk} + G_{ijkl}^{s}\delta\itens{\unem{U}}_{kl} + G_{ij}^\mathrm{a}\delta\Psi\Big)} = H_{ij}(F)-H_{ij}(I)
\end{equation}
can be generally true only if $G_{ijkl}^{s}=0$ and $G_{ij}^\mathrm{a}=0$. Hence, even adding a state-dependent factor $G_{ijkl}^{s}$ and including rotation and isotropic scaling does not resolve the general path-dependence of the cumulative pure shear.

Since any kind of two-dimensional material can be triangulated, path-dependence of the pure shear holds independent of our-triangle-based approach. It is a mere consequence of integrating the instantaneous deformation rate $\itens{V}_{ij}$, which is substantially different from defining deformation with respect to a fixed reference state as usually done in classical elasticity theory \cite{Landau1970}.

\section{Analysis of experimental data}
\label{app:analysisExperimentalData}
\subsection{Quantification of spatially averaged cellular deformation contributions}
\label{app:quantificationSpatialAvgerages}
The equations derived in \ssref{sec:triangleDeformationAndShape}, \ref{sec:largeScaleDeformation}, and \ref{sec:allTogether} hold exactly only for infinitesimal deformations and time intervals. However, experimental data always has a finite acquisition frequency. Here, we discuss how we adapt our theoretical concepts to deal with finite time intervals in practice.

We start from a series $O^k$ of observed states of a cellular network with $k=1,\dots,N_\mathrm{states}$. Each of these states defines cell center positions and cell neighborship relations. The states are registered at times $t^k$, respectively. We denote the corresponding time intervals by $\Delta t^k=t^{k+1}-t^{k}$. As a first step, each of the cellular network states is triangulated according to \sref{sec:triangulation}.

\begin{figure}
  \centering
  \includegraphics{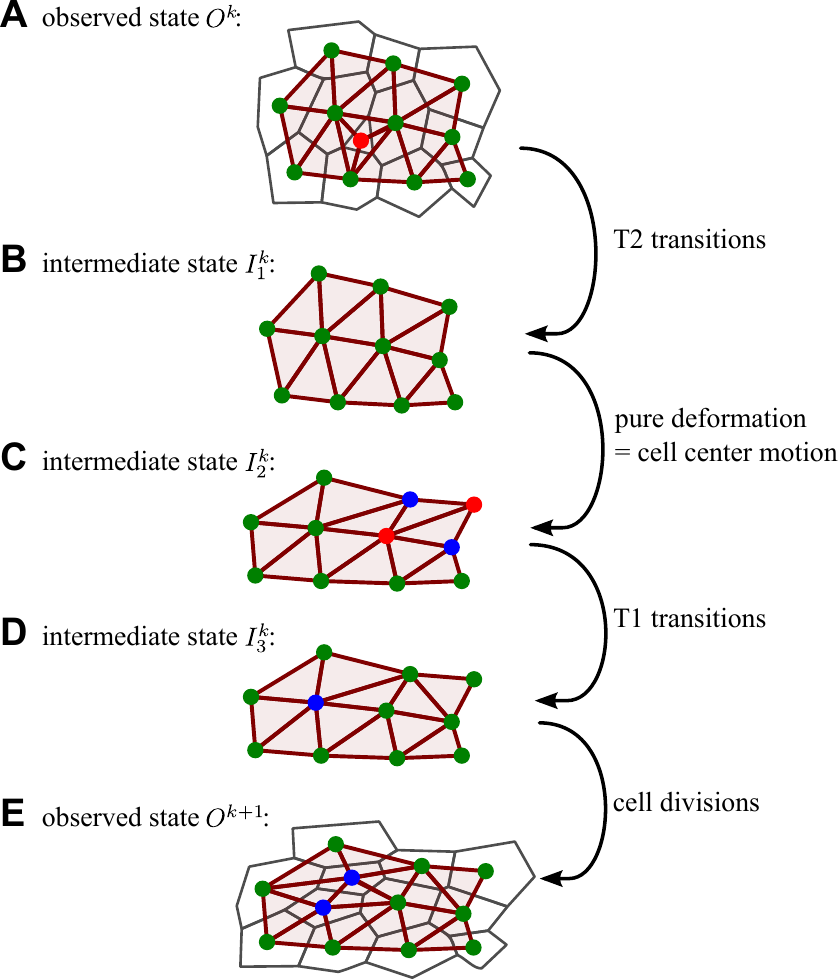}
  \caption{Illustration of the virtual intermediate states $I^k_1$, $I^k_2$, and $I^k_3$ introduced between two observed states $O^k$ and $O^{k+1}$.\label{fig:intermediateStates}}
\end{figure}
To quantify the deformation rate and all cellular contributions to it between two observed states $O^k$ and $O^{k+1}$, we introduce three virtual intermediate network states $I^k_1$, $I^k_2$, and $I^k_3$ (\fref{fig:intermediateStates}, \citep{Merkel2014b,Etournay2015}). By introducing these intermediate states, we shift all topological transitions to the beginning or to the end of the time interval $\Delta t^k$. This separates topological transitions from cell center motions, which now only occur between the states $I^k_1$ and $I^k_2$ (\fref{fig:intermediateStates}B,C). We justify this by the fact that given only the observed data, it is in principle impossible to know at what exact time between $t^k$ and $t^{k+1}$ a given topological transition occurred.  

We define the three intermediate states $I^k_1$, $I^k_2$, and $I^k_3$ based on the observed states $O^k$ and $O^{k+1}$ as follows. 
\begin{enumerate}
  \item The intermediate state $I^k_3$ is defined based on $O^{k+1}$ by reverting all divisions that occur between the observed states $O^k$ and $O^{k+1}$. To this end, each pair of daughter cell centers is fused into a mother cell center. The position of the mother cell center is defined by the average position of the daughter cell centers.
  \item The intermediate state $I^k_1$ is defined based on $O^{k}$ by removing the centers of all cells that undergo a T2 transition between the observed states $O^k$ and $O^{k+1}$.
  \item The intermediate states $I^k_1$ and $I^k_3$ contain the same set of cell centers, which however differ in their positions. Also, the topology of both states is different. We thus define the intermediate state $I^k_2$ based on $I^k_1$ by moving all cell centers to their respective positions in $I^k_3$.
\end{enumerate}
Note that the intermediate states carry just enough information to define the triangulation. Vertex positions, which would be needed to define cellular networks are not contained.

For the precise explanation of how we compute the cellular contributions to the deformation rate, we focus on the pure shear part. Contributions to the isotropic expansion rate or the rotation rate can be computed analogously. In the following, we denote the large-scale shear rate quantified from experimental data and contributions to it with the superscript ``$\mathrm{exp}$''.

We define the pure shear induced by a given kind of topological transition as the negative change of average elongation that is associated with the respective state change (\fref{fig:intermediateStates}). We thus compute the shear rates by T1 transitions $\itens{\nem{T}}_{ij}^\mathrm{exp}$, cell divisions $\itens{\nem{C}}_{ij}^\mathrm{exp}$, and T2 transitions $\itens{\nem{E}}_{ij}^\mathrm{exp}$ as follows:
\begin{align}
  \itens{\nem{T}}_{ij}^\mathrm{exp} &= -\frac{1}{\Delta t^k}\Big[\itens{\nem{Q}}_{ij}(I^k_3)-\itens{\nem{Q}}_{ij}(I^k_2)\Big] \\
  \itens{\nem{C}}_{ij}^\mathrm{exp} &= -\frac{1}{\Delta t^k}\Big[\itens{\nem{Q}}_{ij}(O^{k+1})-\itens{\nem{Q}}_{ij}(I^k_3)\Big]\\
  \itens{\nem{E}}_{ij}^\mathrm{exp} &= -\frac{1}{\Delta t^k}\Big[\itens{\nem{Q}}_{ij}(I^k_1)-\itens{\nem{Q}}_{ij}(O^k)\Big] \text{.}
\end{align}
Here, $\itens{\nem{Q}}_{ij}(X)$ denotes the average triangle elongation in the virtual or observed state $X$. We divide by the time interval $\Delta t^k$ to obtain the respective rate of pure shear.

To compute the large-scale shear rate $\itens{\unem{V}}_{ij}^\mathrm{exp}$, the corotational term $\itens{\nem{J}}_{ij}^\mathrm{exp}$, and the correlation term $\itens{\nem{D}}_{ij}^\mathrm{exp}$, we proceed as follows.
We realized that direct application of \eref{eq:decompositionPureShear} led to large deviations for the fly wing data, which is exact only to first order in the time interval $\Delta t^k$. We thus split the time interval $\Delta t^k$ into $N$ subintervals and then compute $\itens{\unem{V}}_{ij}^\mathrm{exp}$, $\itens{\nem{J}}_{ij}^\mathrm{exp}$, and $\itens{\nem{D}}_{ij}^\mathrm{exp}$ by summing the respective subinterval contributions. To this end, we introduce intermediate states $S^r$ with $r=1,\dots,N-1$, which are defined by interpolating all cell center positions linearly between the states $S^0=I^k_1$ and $S^N=I^k_2$.
Then, the velocity gradient tensor $\itens{V}_{ij}^\mathrm{exp}$ is computed by summing the deformation gradients defined by \eref{eq:UOmegaTrianglesBoundary} for all subintervals:
\begin{equation}
  \itens{V}_{ij}^\mathrm{exp} = \frac{1}{\Delta t^k}\sum_{r=0}^{N-1}{\itens{U}_{ij}^r}
\end{equation}
with the subinterval deformation gradient
\begin{equation}
  \begin{aligned}
    \itens{U}_{ij}^r &= -\frac{\itens{\epsilon}_{im}}{2A(S^r)}\sum_{\alpha=1}^{M}{\Big(\ivec{r}_m^{\alpha+1}(S^r)-\ivec{r}_m^{\alpha}(S^r)\Big)}\times\\
    &\quad\Big(\ivec{r}_j^{\alpha+1}(S^{r+1})+\ivec{r}_j^{\alpha}(S^{r+1})-\ivec{r}_j^{\alpha+1}(S^r)-\ivec{r}_j^{\alpha}(S^r)\Big)
    \text{.}
  \end{aligned}
\end{equation}
Here, $A(S^r)$ and $\vec{r}^{\alpha}(S^r)$ are total triangulation area and position of the center of cell $\alpha$ in state $S^r$, respectively. The inner sum runs over all margin cells $\alpha$ in counter-clockwise order.
The shear rate $\itens{\unem{V}}_{ij}$ is the symmetric, traceless part of $\itens{V}_{ij}$.

The corotational term $\itens{\nem{J}}_{ij}^\mathrm{exp}$ is computed as follows:
\begin{equation}
  \begin{aligned}
    \itens{\nem{J}}_{ij}^\mathrm{exp} &= \sum_{r=0}^{N-1}{\itens{\nem{J}}_{ij}^r}\label{eq:app:corotationalTerm}
  \end{aligned}
\end{equation}
with
\begin{equation}
  \itens{\nem{J}}_{ij}^r = -2\Big[C^r\Psi^r+(1-C^r)(\Phi^{r+1}-\Phi^r)\Big]\itens{\epsilon}_{ik}\itens{\nem{Q}}_{kj}(S^r)\text{.}
\end{equation}
Here, $C^r=\tanh{(2\norm{\itens{\nem{Q}}^r})}/(2\norm{\itens{\nem{Q}}^r})$, where $\itens{\nem{Q}}_{kj}(S^r)$ is the average triangle elongation in state $S^r$, and $\norm{\itens{\nem{Q}}^r}$ and $\Phi^r$ are its norm and angle. The symbol $\Psi^r$ denotes the antisymmetric part of the subinterval deformation tensor $\itens{U}_{ij}^r$, analogous to \eref{eq:decompositionU}.

The correlation term $\itens{\nem{D}}_{ij}^\mathrm{exp}$ is computed as
\begin{equation}
  \begin{aligned}
  \itens{\nem{D}}_{ij}^\mathrm{exp} &= \frac{1}{\Delta t^k}\sum_{r=0}^{N-1}{}\bigg[- \Big(\Big\langle\itens{u}_{kk}^{r}\itens{\nem{q}}_{ij}(S^r)\Big\rangle  -\itens{U}_{kk}^r\itens{\nem{Q}}_{ij}(S^r)\Big) \\
  &\qquad\qquad\qquad\qquad\qquad\qquad\qquad+\Big(\langle\itens{\nem{j}}^{r}_{ij}\rangle - \itens{\nem{J}}_{ij}^r\Big)\bigg]
  \end{aligned}
\end{equation}
Here, $\itens{u}_{kk}^{n,r}$ and $\itens{\nem{j}}^{n,r}_{ij}$ are isotropic expansion and corotational term of triangle $n$ with respect to the subinterval between $S^r$ and $S^{r+1}$, and $\itens{\nem{q}}^n_{ij}(S^r)$ is the elongation of triangle $n$ in state $S^r$. The averaging for a given value of the summation index $r$ is carried out with respect to the triangle areas in state $S^r$.

Finally, we compute the corotational derivative of the average elongation as follows:
\begin{equation}
  \frac{\D\itens{\nem{Q}}_{ij}^\mathrm{exp}}{\D t} = \frac{1}{\Delta t^k}\Big[\itens{\nem{Q}}_{ij}(O^{k+1})-\itens{\nem{Q}}_{ij}(O^{k}) +\itens{\nem{J}}_{ij}^\mathrm{exp}\Big] 
\end{equation}
Here, $\itens{\nem{J}}_{ij}^\mathrm{exp}$ is the corotational term as computed from \eref{eq:app:corotationalTerm}.

Using all these definitions, we can make \eref{eq:decompositionPureShear} hold arbitrarily precise by choosing a sufficiently large value for $N$. For the data shown in \fsref{fig:flyWing} and \ref{fig:flyWing_iso}, we chose $N=100$. Note that this approach to deal with the finiteness of the time intervals $\Delta t^k$ is different from the approaches chosen in our previous publications \citep{Merkel2014b,Etournay2015}.

\subsection{Spatial patterns of shear components}
\label{app:quantificationSpatialPatterns}
To compute spatial patterns of large-scale tissue deformation and their cellular components as in \fref{fig:flyWing_patterns}, we introduce a grid of squared boxes, which are labeled by the index $b$. In \eref{eq:definitionAverage}, we introduced an average over triangles to compute large-scale quantities. Here, we introduce such an average for a given box $b$. For instance, the box-averaged shear rate $\itens{\unem{V}}_{ij}^b=\langle\itens{\unem{v}}_{ij}\rangle_b$ is defined as:
\begin{equation}
  \langle\itens{v}_{ij}\rangle_b = \frac{1}{A_b}\sum_{n}{a^n_b\itens{v}_{ij}^n}\label{eq:box:triangleAverage}
\end{equation}
The sum is over all triangles $n$ that have an overlap with box $b$, and $a^n_b$ is the area of this overlap. The normalization factor $A_b$ is the overlap area between box $b$ and the triangulation, i.e.\ $A_b=\sum_{n}{a^n_b}$.

\subsubsection{Infinitesimal time intervals}
Here and in the following, we focus our discussion on the computation of the pure shear part and its cellular contributions. First we ask how the box-averaged shear rate $\itens{\unem{V}}_{ij}^b$ decomposes into cellular contributions for an infinitesimal time interval $\delta t$ and in the absence of topological transitions. To this end, we insert the relation between single triangle shear rate and triangle shape, \eref{eq:triangleShearRateAndQ}, into \eref{eq:box:triangleAverage} and obtain an equation that is analogous to \eref{eq:decompositionShearRateWithoutTopologicalTransitions}:
\begin{equation}
  \itens{\unem{V}}_{ij}^b = \frac{\D\itens{\nem{Q}}_{ij}^b}{\D t} + \itens{\nem{D}}_{ij}^b\text{.}\label{eq:box:shearRateNoTt}
\end{equation}
However here, the corotational time derivative contains an additional term $\itens{\nem{B}}_{ij}^b$:
\begin{equation}
  \frac{\D\itens{\nem{Q}}_{ij}^b}{\D t} = \frac{\delta\itens{\nem{Q}}_{ij}^b}{\delta t} + \itens{\nem{B}}_{ij}^b + \frac{\delta\itens{\nem{J}}_{ij}^b}{\delta t}
\end{equation}
with the definitions
\begin{align}
  \itens{\nem{Q}}_{ij}^b &= \langle\itens{\nem{q}}_{ij}\rangle_b \\
  \itens{\nem{B}}_{ij}^b &= -\left(\left\langle\itens{\nem{q}}_{ij}\frac{\d}{\d t}(\ln{f_b})\right\rangle_b-  \itens{\nem{Q}}_{ij}^b\left\langle\frac{\d}{\d t}(\ln{f_b})\right\rangle_b\right) \\
  \delta\itens{\nem{J}}_{ij}^b &= -2\Big[C_b\langle\delta\psi\rangle_b+(1-C_b)\delta\Phi_b\Big]\itens{\epsilon}_{ik}\itens{\nem{Q}}_{kj}^b\text{.}
\end{align}
Here, $f^n_b = a^n_b/a^n$ is the area fraction of triangle $n$ that is inside box $b$, and $C_b=\tanh{(2\norm{\itens{\nem{Q}}^b})}/2\norm{\itens{\nem{Q}}^b}$. The symbols $\norm{\itens{\nem{Q}}^b}$ and $\Phi_b$ denote norm and angle of the average elongation tensor $\itens{\nem{Q}}_{ij}^b$, respectively.
The correlation term in \eref{eq:box:shearRateNoTt} is defined by
\begin{equation}
  \itens{\nem{D}}_{ij}^b = - \Big(\big\langle\itens{v}_{kk}\itens{\nem{q}}_{ij}\big\rangle_b-  \langle\itens{v}_{kk}\rangle_b\itens{\nem{Q}}_{ij}^b\Big) + \frac{1}{\delta t}\Big(\langle\delta\itens{\nem{j}}_{ij}\rangle_b - \delta\itens{\nem{J}}_{ij}^b\Big)\text{.}
\end{equation}
\eref{eq:decompositionShearRateWithoutTopologicalTransitions} describes a triangulation that is followed as it moves through space, whereas here, we consider a box $b$ that is fixed in space. Correspondingly, the we interpret the additional term $\itens{\nem{B}}_{ij}^b$ in the corotational derivative as a convective term.

\subsubsection{Finite time intervals}
To practically compute the pure shear contributions for a given box $b$ for experimental image data, we proceed similar to the previous section. We consider again a finite time interval $\Delta t^k$ between two subsequent observed states $O^k$ and $O^{k+1}$. To separate pure shear contributions by topological transitions from contributions by cell center motion, we introduce again the intermediate states illustrated in \fref{fig:intermediateStates}. Correspondingly, the shear rates by T1 transitions $\itens{\nem{T}}_{ij}^{\mathrm{exp},b}$, by cell divisions $\itens{\nem{C}}_{ij}^{\mathrm{exp},b}$, and by T2 transitions $\itens{\nem{E}}_{ij}^{\mathrm{exp},b}$ are defined as:
\begin{align}
  \itens{\nem{T}}_{ij}^{\mathrm{exp},b} &= -\frac{1}{\Delta t^k}\Big[\itens{\nem{Q}}_{ij}^b(I^k_3)-\itens{\nem{Q}}_{ij}^b(I^k_2)\Big] \\
  \itens{\nem{C}}_{ij}^{\mathrm{exp},b} &= -\frac{1}{\Delta t^k}\Big[\itens{\nem{Q}}_{ij}^b(O^{k+1})-\itens{\nem{Q}}_{ij}^b(I^k_3)\Big]\\
  \itens{\nem{E}}_{ij}^{\mathrm{exp},b} &= -\frac{1}{\Delta t^k}\Big[\itens{\nem{Q}}_{ij}^b(I^k_1)-\itens{\nem{Q}}_{ij}(O^k)^b\Big] \text{.}
\end{align}
The tensors $\itens{\nem{Q}}_{ij}^b(X)$ denote the box-averaged triangle elongation in the virtual or observed state $X$.

To compute the box-averaged shear rate $\itens{\unem{V}}_{ij}^{\mathrm{exp},b}$, the convective term $\itens{\nem{B}}_{ij}^{\mathrm{exp},b}$, the corotational term $\itens{\nem{J}}_{ij}^{\mathrm{exp},b}$, and the correlations $\itens{\nem{D}}_{ij}^{\mathrm{exp},b}$ between $O^k$ and $O^{k+1}$, we use the subintervals and the states $S^r$ with $r=0,\dots,N$ introduced in the previous section. We again compute the quantities for each subinterval separately and then sum over the subintervals:
\begin{align}
  \itens{\unem{V}}_{ij}^{\mathrm{exp},b} &= \frac{1}{\Delta t^k}\sum_{r=0}^{N-1}{\langle\itens{\unem{u}}_{ij}^{r}\rangle} \\
  \itens{\nem{B}}_{ij}^{\mathrm{exp},b} &= -\frac{1}{\Delta t^k}\sum_{r=0}^{N-1}{} \bigg(\left\langle\itens{\nem{q}}_{ij}(S^r)\frac{\Delta f^{r}_b}{f^{r}_b}\right\rangle_b\nonumber\\
  &\qquad\qquad\qquad\qquad-\itens{\nem{Q}}_{ij}^b(S^r)\left\langle\frac{\Delta f^{r}_b}{f^{r}_b}\right\rangle_b\bigg) \\
  \itens{\nem{J}}_{ij}^{\mathrm{exp},b} &= \sum_{r=0}^{N-1}{\itens{\nem{J}}_{ij}^{b,r}}\\
  \itens{\nem{J}}_{ij}^{b,r} &= -2\Big[C^{b,r}\Psi^{b,r}+(1-C^{b,r})(\Phi^{b,r+1}-\Phi^{b,r})\Big]\times\nonumber\\
  &\qquad\qquad\qquad\qquad\qquad\qquad\qquad\qquad\itens{\epsilon}_{ik}\itens{\nem{Q}}_{kj}^b(S^r)\\
  \itens{\nem{D}}_{ij}^{\mathrm{exp},b} &= \frac{1}{\Delta t^k}\sum_{r=0}^{N-1}{}\bigg[-\Big(\big\langle\itens{u}_{kk}^{r}\itens{\nem{q}}_{ij}(S^r)\big\rangle_b-  \langle\itens{u}_{kk}^{r}\rangle_b\itens{\nem{Q}}_{ij}^b(S^r)\Big) \nonumber\\
  &\qquad\qquad\qquad\qquad\qquad\qquad +\Big(\langle\itens{\nem{j}}^{r}_{ij}\rangle_b - \itens{\nem{J}}_{ij}^{b,r}\Big)\bigg]\text{.}
\end{align}
Here, $\itens{u}_{kk}^{n,r}$ and $\itens{\unem{u}}_{ij}^{n,r}$ are trace and symmetric, traceless part of the deformation tensor of triangle $n$ according to \eref{eq:triangleMu} with respect to the subinterval between $S^r$ and $S^{r+1}$,
$\itens{\nem{q}}^n_{ij}(S^r)$ is the elongation of triangle $n$ in state $S^r$, $f^{n,r}_b$ is the value of $f^n_b$ in state $S^r$, and its change is $\Delta f^{n,r}_b=f^{n,r+1}_b-f^{n,r}_b$.
We furthermore used $C^{b,r}=\tanh{(2\norm{\itens{\nem{Q}}^{b,r}})}/(2\norm{\itens{\nem{Q}}^{b,r}})$, where $\norm{\itens{\nem{Q}}^{b,r}}$ and $\Phi^{b,r}$ are norm and angle of the box-averaged elongation in state $S^r$, $\itens{\nem{Q}}^b_{ij}(S^r)$. The symbol $\Psi^{b,r}$ denotes the antisymmetric part of the box-averaged deformation tensor in state $r$ and the tensor $\itens{\nem{j}}^{n,r}_{ij}$ denotes the corotational term for triangle $n$ with respect to the subinterval between $S^r$ and $S^{r+1}$.
Finally, the corotational derivative of the box-averaged elongation is computed as
\begin{equation}
  \frac{\D\itens{\nem{Q}}_{ij}^{\mathrm{exp},b}}{\D t} = \frac{1}{\Delta t^k}\Big[\itens{\nem{Q}}_{ij}^b(O^{k+1})-\itens{\nem{Q}}_{ij}^b(O^{k}) +\itens{\nem{J}}_{ij}^{\mathrm{exp},b}\Big]\text{.}
\end{equation}
For the patterns shown in \fref{fig:flyWing_patterns}, we used $N=100$ subintervals.

\label{app:lastAppendix}

\printbib
\end{document}